\documentclass[fleqn,usenatbib]{mnras}

\usepackage{newtxtext,newtxmath}

\usepackage[T1]{fontenc}
\usepackage{ae,aecompl}
\usepackage{longtable}

\usepackage{graphicx}	
\usepackage{amsmath}	

\newcommand{\aifa}{Argelander-Institut f\"{u}r Astronomie, Universit\"{a}t Bonn, Auf dem H\"{u}gel 71, 53121, Bonn, Germany}
\newcommand{\ari}{Astrophysics Research Institute, Liverpool John Moores University, 146 Brownlow Hill, Liverpool L3 5RF, UK}
\newcommand{\uconn}{Department of Physics, University of Connecticut, 196A Auditorium Road, Storrs, CT 06269, USA}
\newcommand{\kiaa}{Kavli Institute for Astronomy and Astrophysics, Peking University, Beijing 100871, China}
\newcommand{\pkudoa}{Department of Astronomy, School of Physics, Peking University, Beijing 100871, China}
\newcommand{\shao}{Shanghai Astronomical Observatory, Chinese Academy of Sciences, 80 Nandan Road, Shanghai 200030, People's Republic of China}
\newcommand{\naoj}{National Astronomical Observatory of Japan, 2-21-1 Osawa, Mitaka, Tokyo 181-8588, Japan}
\newcommand{\ku}{Department of Physics and Astronomy, University of Kansas, 1251 Wescoe Hall Dr., Lawrence, KS 66045, USA}
\newcommand{\aride}{Astronomisches Rechen-Institut, Zentrum f{\"u}r Astronomie der Universit{\"a}t Heidelberg, M{\"o}nchhofstra{\ss}e 12-14, D-69120 Heidelberg, Germany}
\newcommand{\udc}{Departamento de Astronom{\'i}a, Universidad de Chile, Camino el Observatorio 1515, Las Condes, Santiago, Chile}
\newcommand{\jao}{Joint ALMA Observatory, Alonso de C{\'o}rdova 3107, Vitacura, Santiago, Chile}
\newcommand{\uof}{Department of Astronomy, University of Florida, PO Box 112055, USA}
\newcommand{\mpia}{Max Planck Institute for Astronomy, K{\"o}nigstuhl 17, D-69117 Heidelberg, Germany}
\newcommand{\smith}{Center for Astrophysics $\vert$ Harvard \& Smithsonian, 60 Garden Street, Cambridge, MA 02138, USA}
\newcommand{\jbca}{UK ALMA Regional Centre Node, Jodrell Bank Centre for Astrophysics, The University of Manchester, Manchester M13 9PL, UK}

\title[YMC formation in the Galactic centre]{The initial conditions for young massive cluster formation in the Galactic Centre: convergence of large-scale gas flows}

\author[B. A. Williams et al.]{Bethan A. Williams$^{1}$\thanks{E-mail: B.A.Williams@2015.ljmu.ac.uk (BAW)},
Daniel L. Walker$^{2,3}$,
Steven N. Longmore$^{1}$,
A.~T.~Barnes$^{4}$,
\newauthor
Cara Battersby$^{2}$,
Guido Garay$^{5}$,
Adam Ginsburg$^{6}$,
Laura Gomez$^{7}$,
\newauthor
Jonathan D. Henshaw$^{8}$,
Luis C. Ho$^{9,10}$,
J.~M.~Diederik~Kruijssen$^{11}$,
Xing Lu$^{12,13}$,
\newauthor
Elisabeth A. C. Mills$^{14}$,
Maya A. Petkova$^{11}$
and Qizhou Zhang$^{15}$
\\
$^{1}$\ari \\
$^{2}$\uconn \\
$^{3}$\jbca \\
$^{4}$\aifa \\
$^{5}$\udc \\
$^{6}$\uof \\
$^{7}$\jao \\
$^{8}$\mpia \\
$^{9}$\pkudoa \\
$^{10}$\kiaa \\
$^{11}$\aride \\
$^{12}$\naoj \\
$^{13}$\shao \\
$^{14}$\ku \\
$^{15}$\smith
}

\date{Accepted XXX. Received YYY; in original form ZZZ}

\pubyear{2022}

\begin{document}
\label{firstpage}
\pagerange{\pageref{firstpage}--\pageref{lastpage}}
\maketitle

\begin{abstract}
Young massive clusters (YMCs) are compact ($\lesssim$1 pc), high-mass (>10${}^4$ M${}_{\odot}$) stellar systems of significant scientific interest. Due to their rarity and rapid formation, we have very few examples of YMC progenitor gas clouds before star formation has begun. As a result, the initial conditions required for YMC formation are uncertain. We present high-resolution (0.13$^{\prime\prime}$, $\sim$1000\,au) ALMA observations and Mopra single-dish data, showing that Galactic Centre dust ridge `Cloud d' (G0.412$+$0.052, mass~$=7.6 \times 10^4$\,M$_{\odot}$, radius~$=3.2$\,pc) has the potential to become an Arches-like YMC (10$^4$ M$_{\odot}$, r~$\sim$\,1\,pc), but is not yet forming stars. This would mean it is the youngest known pre-star forming massive cluster and therefore could be an ideal laboratory for studying the initial conditions of YMC formation. We find 96 sources in the dust continuum, with masses $\lesssim$3\,M$_{\odot}$ and radii of $\sim$10${}^3$\,au. The source masses and separations are more consistent with thermal rather than turbulent fragmentation. It is not possible to unambiguously determine the dynamical state of most of the sources, as the uncertainty on virial parameter estimates is large. We find evidence for large-scale ($\sim$1 pc) converging gas flows, which could cause the cloud to grow rapidly, gaining 10$^4$\,M$_{\odot}$ within 10$^5$\,yr. The highest density gas is found at the convergent point of the large-scale flows. We expect this cloud to form many high-mass stars, but find no high-mass starless cores. If the sources represent the initial conditions for star formation, the resulting IMF will be bottom-heavy.
\end{abstract}






\section{Introduction}

Young massive clusters (YMCs) are gravitationally bound stellar systems with masses $\gtrsim$ 10${}^4$\,M${}_{\odot}$, radii $\sim$1\,pc, and ages $\lesssim$\,100 Myr \citep{portegies10}. The large number of co-eval stars within YMCs provide an important astrophysical laboratory to study the stellar initial mass function, stellar evolution and stellar dynamics. As local-Universe analogues of young globular clusters  \citep{elmegreen97,kruijssen15b,pfeffer18}, studying nearby YMCs provides an important way to understand the formation and early evolution of stars and clusters in extreme environments across cosmological timescales.

Despite their importance, we still have limited observational examples of YMC progenitor clouds before star formation has begun \citep{ginsburg12, longmore12, urquhart13, contreras17, jackson18}. Two main YMC formation mechanisms have been proposed -- a monolithic ``in situ" mode and a hierarchical ``conveyor belt" mode \citep[see, e.g.][for a review]{longmore14}. In the monolithic scenario, all gas is contained within the final cluster volume before star formation begins. After forming its stars, the remaining gas is lost from the cluster, decreasing the global gravitational potential and the cluster expands towards its final, unembedded phase. In the hierarchical scenario, gas is initially more extended than the final cluster volume, with both the extended gas cloud and the embedded protostellar population undergoing global gravitational collapse simultaneously. Studies of young cluster and progenitor cloud populations show that the latter ``conveyor belt" formation mode, where gas accretion and star formation occur simultaneously, better reproduces their observed properties \citep{longmore14, walker15, krumholz20}. 

Observationally differentiating between these formation mechanisms is, however, difficult. A prediction of the hierarchical conveyor belt mode of YMC formation is that there should exist gas clouds with mass of 10$^5$\,M$_\odot$ and radii of a few pc, which contain a small amount of star formation activity. Without this on-going star formation it can not be determined if a quiescent cloud will simply collapse to form a very high density proto-cluster in future (i.e. in situ). Identifying massive molecular clouds on the cusp of forming stars then provides the rare opportunity to observe the very initial stages of these YMC formation mechanisms, and provides insight to the dynamics of the cloud prior to the formation of stars.

Despite extensive observational searches \citep[e.g.][]{bressert12,ginsburg12, urquhart13, longmore13, longmore17}, such clouds have remained elusive in the Milky Way. The most promising examples to date have generally been found in the `Central Molecular Zone' (CMZ) -- the inner few hundred pc of the Galaxy \citep{henshaw22b} (the Henshaw+22 reference has been added in this edit). In particular, a region of the CMZ known as the `dust ridge' \citep{lis94}, contains a collection of six massive (10${}^5$\,M${}_{\odot}$), compact (radius~$\sim$1-3\,pc), and largely quiescent clouds \citep[excluding the Sagittarius B2 complex, one of the most active sites of high-mass star formation in the entire Milky Way, e.g.][]{ginsburg18,ginsburg18b,schworer19} orbiting at $\sim$100\,pc from the Galactic Centre \citep{kruijssen15,kdlplus19,dkl19,petkova21} which have been identified as potential progenitors to YMCs \citep[e.g.][]{longmore13b,rathborne15,walker15, barnes19}.

A subset of the dust ridge clouds have been studied in detail, from pc scales down to the scale of individual cores ($\sim$1000\,au)
\citep[e.g.][]{lis94, immer12, longmore12, longmore13b, rodriguez13, rathborne14, rathborne15, mills15, walker18, walker21, barnes19, lu19b, lu19, lu20, lu21, henshaw19, henshaw22a, battersby20, hatchfield20}. Based on the evolution of dense gas structure and analysis of the gas kinematics, \cite{walker15, walker16} and \cite{barnes19} conclude that YMCs forming from these clouds are more likely to do so in a way that is more consistent with the predictions of a hierarchical conveyor belt mode. Intriguingly, \cite{henshaw16c} observed a regular, corrugated velocity field -- which they referred to as `wiggles' -- within the same contiguous gas stream as the dust ridge clouds, located $\sim$20 pc upstream from the dust ridge in projection. They found that the velocity extremes correlate with regularly spaced ($\sim$8\,pc) massive, compact molecular clouds. They interpreted the velocity wiggles as kinematic evidence of cloud formation via large-scale gravitational collapse. If this interpretation is correct, the dust ridge clouds are potentially more evolved, collapsing `wiggles', providing a key laboratory for studying YMC formation.

Perhaps unsurprisingly given their important role for understanding star formation in extreme environments, the Galactic Centre gas clouds with signs of ongoing star formation activity have been studied in the most detail \citep[e.g.][]{lu19b,lu20,walker21}. Unfortunately -- at least as far as searching for a pre-star-forming YMC progenitor cloud is concerned -- even the most massive and previously most quiescent of these, G0.253$+$0.016 (the `Brick'), has now been shown to be unambiguously forming stars \citep{walker21,henshaw22a}. \cite{walker21} find that this is only a small grouping of 18 low-to-intermediate mass sources, and is contained to a small area rather than widespread throughout the `Brick'. \cite{henshaw22a} find that the Brick may have already formed a small ($\sim$10${}^3$\,M${}_{\odot}$) cluster.

In the ongoing search for a truly pre-star-forming YMC progenitor cloud, we therefore turn our attention to the least studied of the dust ridge clouds, G0.412$+$0.052 (hereafter referred to as cloud `d'). Despite having a similar mass and radius to other dust ridge clouds, cloud `d' shows no signs of star formation  on $\gtrsim$0.01\,pc scales \citep{walker18,barnes19}. In this study, we present high angular resolution (0.13${}^{\prime\prime}$) ($\sim$1000\,au) ALMA Band 6 observations towards the peak of the single-dish continuum emission in cloud `d' \citep[clump `d6' in][]{walker18} using the same observational and spectral setup as \citet{walker21} who found embedded star formation on 1000\,au scales in the `Brick'. We aim to determine whether or not star formation is occurring at the scale of individual cores and understand the fate of this cloud by investigating the gas density distribution in relation to large scale gas kinematics.

\section{Observations and Data}

\subsection{Observations}

\begin{figure*}
		\begin{center}
		\includegraphics[width=\textwidth]{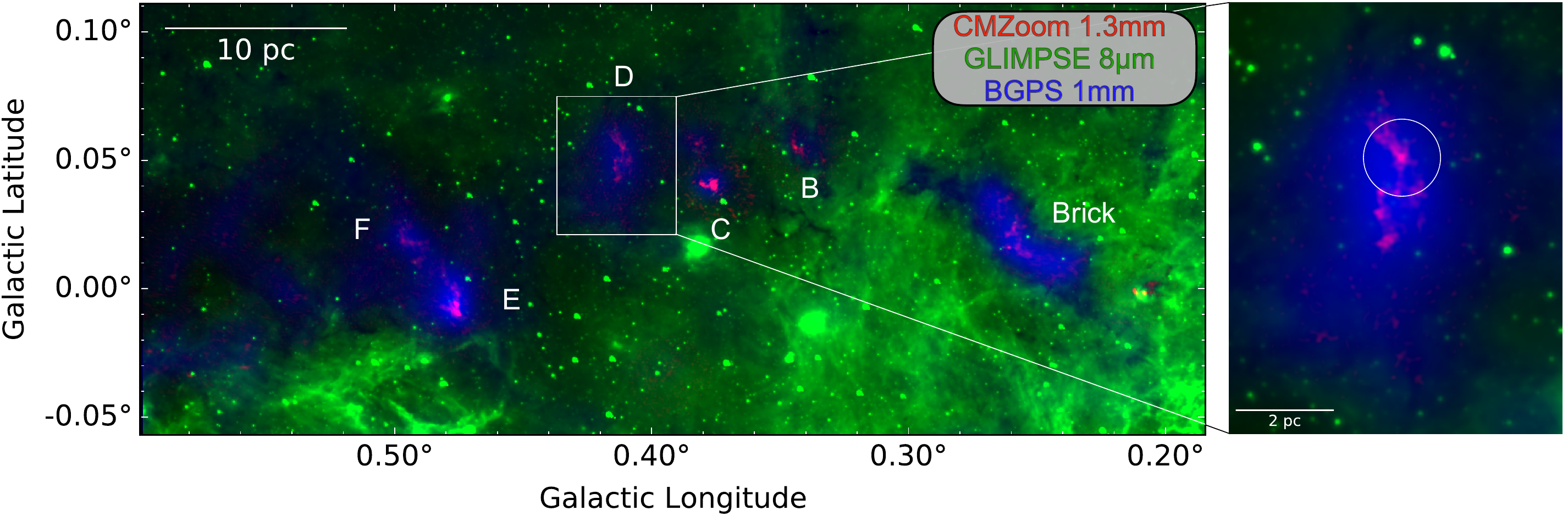}
		\caption{\textbf{Left}: Three-colour image of the Galactic centre dust ridge. {\emph{Red}}: SMA 1.3~mm dust continuum \citep{battersby20}, {\emph{Green}}: Spitzer/GLIMPSE 8~$\mu$m emission \citep{churchwell09}, {\emph{Blue}}: Bolocam Galactic Plane Survey 1~mm dust continuum \citep[BGPS,][]{ginsburg13}. \textbf{Right}: Zoom-in on dust ridge cloud `d' (G0.412$+$0.052) The white circle corresponds to the primary beam field of view of the ALMA observation towards source `d6' \citep{walker18} reported in this paper.}
		\label{fig:overview}
		\end{center}
\end{figure*}

We obtained single-pointing (Table \ref{tab:obsparam}), high-sensitivity and high-angular-resolution dust continuum and molecular line observations towards clump `d6'  in cloud `d' with ALMA \citep[R.A. (J2000) 17:46:23.0, Dec. (J2000) -28:33:23.5;][see Figure \ref{fig:overview}]{walker18}. The observations were performed during ALMA's Cycle 4 (project ID: 2016.1.00949.S, PI: D. Walker), at a frequency of $\sim$230GHz (1.3 mm, frequency band 6). The angular resolution is $\sim$0.13${}^{\prime\prime}$, which allows us to resolve scales of 1000\,au, the scale of individual cores, at a distance of 8.1 kpc \citep{abuter19}. The observations contain 7 spectral windows, 5 of which targeted specific molecular transitions (Table \ref{tab:specparam}) in the lower sideband with a spectral resolution of $\sim$0.77\,km\,s$^{-1}$. The remaining two spectral windows were dedicated to broad-band continuum detection in the upper sideband, with a spectral resolution of $\sim$2.5 km\,s$^{-1}$. The total aggregate bandwidth is approximately 5.6 GHz. The project was observed across 4 individual execution blocks between October 2016 and April 2017. Each execution used 41-45 antennas, with baselines ranging from 15 - 3696 m. Full observation parameters and spectral setup details can be found in Tables \ref{tab:obsparam} \& \ref{tab:specparam} respectively.


\begin{table*}
	\centering
	\caption{Details of the four observed execution blocks. Listed are the observation dates, nominal array configurations, number of 12 m antennas in the array, full range of antenna baseline lengths, total time on source, and the bandpass, flux, and phase calibrators used for each observation.}
	\label{tab:obsparam}
	\begin{tabular}{ccccccccc} 
		\hline
		Date & Array & Antennas & Baselines & Time on source & Bandpass & Flux & Phase\\
		(d/m/y) & configuration & \# & (m) & (minutes) & calibrator & calibrator & calibrator\\
		\hline
		14/10/2016 & C40-6 & 41 & 18 - 2535 & 45.37 & J1924-2914 & J1924-2914 & J1744-3116\\
		25/04/2017 & C40-3 & 41 & 15 - 450 & 27.22 & J1924-2914 & J1924-2914 & J1744-3116\\
		19/07/2017 & C40-6 & 42 & 18 - 3696 & 45.37 & J1924-2914 & J1733-1304 & J1744-3116\\
		08/08/2017 & C40-6 & 45 & 21 - 3696 & 45.37 & J1924-2914 & J1733-1304 & J1744-3116\\
		\hline
		\hline
	\end{tabular}
\end{table*}

\begin{table}
	\centering
	\caption{Overview of the spectral setup used for our ALMA observation. The specific line(s) targeted per spectral window are given, along with the corresponding central frequency ($\nu_{\rm cent}$), bandwidth (BW), and spectral resolution in terms of velocity ($\Delta\nu$).}
	\label{tab:specparam}
	\begin{tabular}{cccc} 
		\hline
		Spectral & $\nu_{\rm cent}$ & BW & $\Delta\nu$\\
		window & (GHz) & (GHz) & (km s$^{-1}$)\\
		\hline
		SiO (5-4) & 217.105 & 0.234 & 0.78\\
		H$_2$CO (3$_{0,3}$ - 2$_{0,2}$) & 218.222 & 0.234 & 0.78\\
		H$_2$CO (3$_{2,2}$ - 2$_{2,1}$) & 218.476 & 0.234 & 0.78\\
		H$_2$CO (3$_{2,1}$ - 2$_{2,0}$) & 218.760 & 0.234 & 0.78\\
		$^{13}$CO (2-1)/CH$_3$CN (12-11) & 220.709 & 0.934 & 0.77\\
		Continuum & 232.500 & 1.875 & 2.50\\
		Continuum & 235.000 & 1.875 & 2.47\\
		\hline
		\hline
	\end{tabular}
\end{table}

\subsection{Image Cleaning \& Processing}

The ALMA pipeline calibrated data sets for each execution block were combined to obtain final data products, which were then imaged in CASA Version 5.6.0.68 \citep{mcmullin07}.

Prior to generating the dust continuum, any channels with spectral line contamination were flagged. No line emission is detected in the broad spectral windows in the upper sideband, and only a small fraction of the channels in the other spectral windows contain line emission. We estimate that no more than 10\% of the aggregate bandwidth is flagged due to line contamination, and the effective bandwidth used for continuum generation is $\sim$5~GHz.

We use CASA's {\tt tclean} task to image both the continuum and line cubes. Due to the complex structure of cloud `d', we opt for {\tt tclean}'s `automasking' mode over 10${}^5$ iterations. A cleaning threshold of 15\,$\mu$Jy (the rms sensitivity of the continuum image) was used, with Briggs weighting and a {\tt robust} parameter of 0.5. The resultant image has a synthesised beam size of $0.14{}^{\prime\prime}\times 0.11^{\prime\prime}$ ($\sim$1100\,au $\times$ 890\,au).

The data suffer with an inherent limited flux recovery due to the incomplete \textit{uv}-coverage of the interferometer. We account for this by combining our interferometric data with the most appropriate large-scale data. To recover the continuum emission, we use $\sim$1$^{\prime\prime}$ resolution data from \cite{barnes19}, which combines observations from ALMA's 12m array and 7m array with BOLOCAM Galactic Plane Survey \citep[BGPS, ][]{ginsburg13} data. As their observations were made at a different frequency to the BGPS data, the data had to be scaled in order to be combined. \cite{barnes19} scale the BGPS data to their 12m$+$7m data using the relation:

\begin{equation}\label{ash}
    \frac{F_\mathrm{ALMA}}{F_\mathrm{BOLOCAM}} = \Bigg(\frac{\nu_\mathrm{ALMA}}{\nu_\mathrm{BOLOCAM}}\Bigg)^{\alpha_{\nu}},
\end{equation}

\noindent where $F$ (Jy\,beam${}^{-1}$) and $\nu$ (GHz) are the continuum intensities and approximate central frequencies of the \cite{barnes19} ALMA observations and the BOLOCAM observations, and $\alpha_{\nu}$ is the spectral index, quantifying how the intensity of dust emission varies with frequency.

Our ALMA data was taken at a frequency of $\sim$225\,GHz, while the \cite{barnes19} data has a frequency of $\sim$259\,GHz, meaning that we have to use the same scaling relationship before combining the two data sets. Using $\alpha_{\nu} = 3.75$, as \cite{barnes19} did, we get:

\begin{equation}\label{ash2}
    \frac{F_\mathrm{ALMA,225GHz}}{F_\mathrm{ALMA,259GHz}} = \Bigg(\frac{\nu_\mathrm{ALMA,225GHz}}{\nu_\mathrm{ALMA,259GHz}}\Bigg)^{\alpha_{\nu}} \approx \Bigg(\frac{225}{259}\Bigg)^{3.75} \approx 0.6.
\end{equation}

\noindent We use the {\tt feather} task in CASA when combining the \cite{barnes19} data with the cleaned ALMA data. The total flux of the image before feathering is 0.0247 Jy, whereas after feathering it is 1.077 Jy. Any discussion regarding the dust continuum hereafter refers to results obtained using fully combined maps of our cleaned ALMA data and that within \cite{barnes19}. 

We note that some of our initial ALMA data is lost due to the different footprints of the feathered data sets. We do not, however, lose any of the main area of dust emission, and so continue with this feathered image.

After feathering, the rms continuum sensitivity is $\sim$15\,$\mu$Jy, corresponding to a 5$\sigma$ mass sensitivity of $\sim$0.1\,M$_{\odot}$, assuming a dust temperature of 20\,K (see $\S$~\ref{sub:cont_source_props}).

We split out the target spectral windows using CASA's {\tt split} task and defining the spectral windows that are our desired targets. We subtract the continuum from the isolated lines using the {\tt uvcontsub} task. We use the {\tt tclean} task to image the lines. We use the interactive manual clean so we can stop defining masks when all the emission had been cleaned and the residuals look like noise with no remaining structure. We again use Briggs weighting and a {\tt robust} parameter of 0.5, although here we use a cleaning threshold of 0.06\,mJy. Finally, we use the {\tt immoments} task to produce the desired moment maps.

For the spectral lines, we only have 12m data, and so we are missing the larger scale structure. We use CASA's {\tt imsmooth} task to perform a spatial Gaussian smoothing on the molecular line data to improve the signal to noise. We set the major and minor axes parameters of the Gaussian smoothing kernel to equal $\sim$0.34${}^{\prime\prime}$, double the size of the major axis of the synthesised beam. This increased the rms of the line images from $\sim$0.6\,mJy to $\sim$1\,mJy. Any discussion regarding molecular lines hereafter refers to results obtained from this Gaussian smoothing. The resulting moment maps of the lines following this smoothing can be found in the supplementary online material. We will discuss the morphology and kinematics of this emission later in the paper.

\section{Results}
\label{sec:results}

\subsection{Spatial distribution, mass, radius and density of compact continuum sources}
\label{sub:cont_source_props}

\begin{figure*}
	    \begin{center}
        \includegraphics[width=18cm]{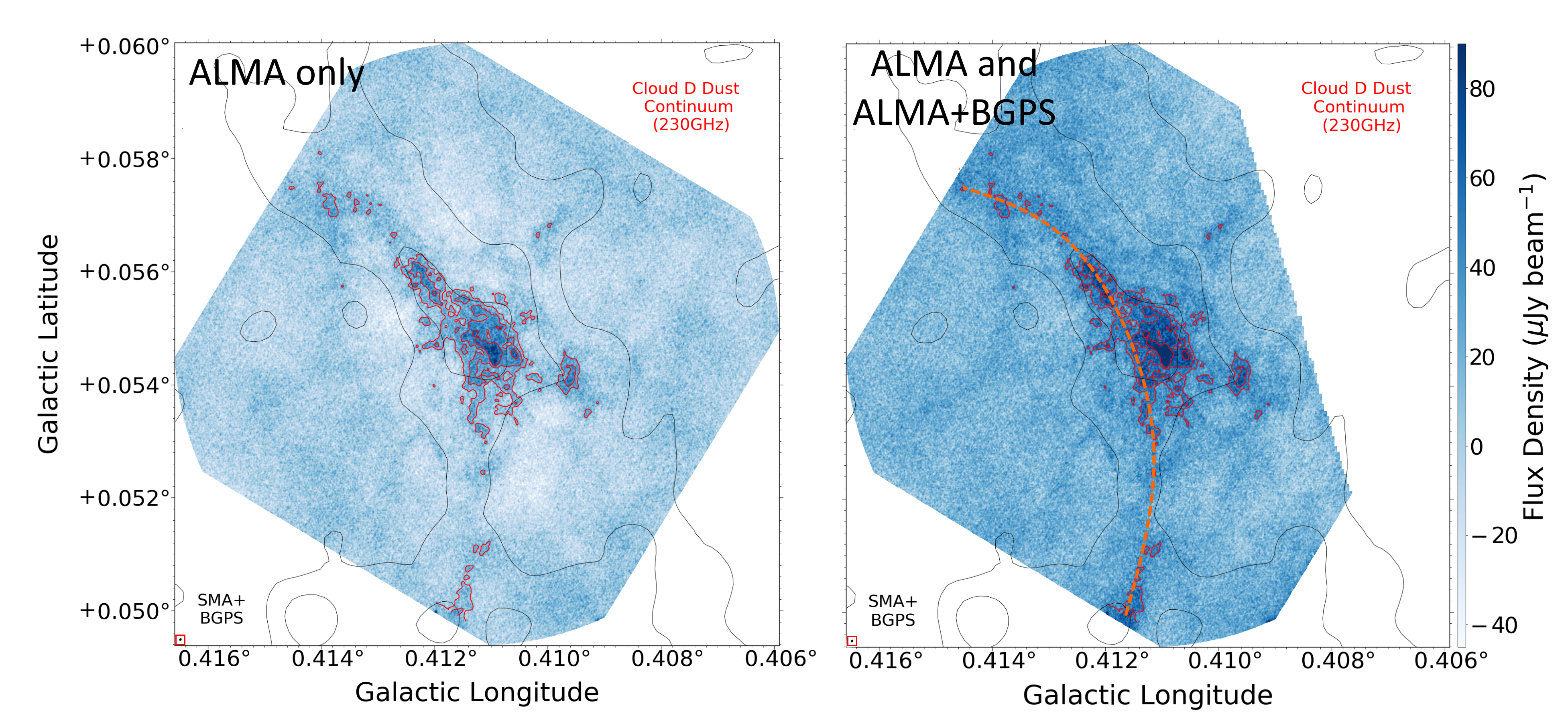}
		\caption{Left: Dust continuum map of our ALMA data only. Red contours show structure in the continuum data at levels of 45, 60 and 80 $\mu$Jy (levels of 3, 4 and 5.3$\sigma$ respectively). Black contours show just the SMA$+$BGPS continuum data from \citet{walker18}. The black circle in the bottom left (highlighted by a red box) represents the synthesised beam. Filamentary structure is not as clear in the unfeathered data. Additionally, white `patches' can be seen surrounding the filament. These `negative bowls' are a sign of missing zero-spacing data (i.e. the large-scale structure has been spatially filtered out). Right: Our ALMA data feathered with ALMA$+$BGPS data \citep{barnes19}. Red contours show structure in the continuum data at levels of 45, 60 and 80 $\mu$Jy. Black contours show just the SMA$+$BGPS continuum data from \citet{walker18}. The black circle in the bottom left (highlighted by a red box) represents the synthesised beam. Curved filamentary structure (represented by a dashed orange line) can be seen with a roughly central mass concentration. A section of the map is missing due to our ALMA data and the \citet{barnes19} ALMA$+$BGPS data having different footprints.}
		\label{fig:contfeath}
		\end{center}
    \end{figure*}

The left-hand panel of Figure \ref{fig:contfeath} displays the ALMA-only 1.3\,mm dust continuum map for cloud `d', focused on `d6' \citep{walker18}, along with overlaid contours of both the feathered continuum data (red) and the combined SMA and Bolocam Galactic Plane Survey (BGPS) data \citep[black;][]{walker18}. The right-hand panel displays the continuum map for our data feathered with ALMA$+$BGPS data \citep{barnes19}, with the same contours as the left-hand panel. Note that a section of the feathered continuum map is missing on the upper right side due to the ALMA data and the ALMA$+$BGPS data having different footprints. However, no obvious structure has been lost in this process (see left-hand panel of Figure \ref{fig:contfeath} for the unfeathered ALMA data with a complete footprint). We detect dense substructure within cloud `d' with a filamentary morphology. The mass concentration peaks in the middle of the image, which corresponds to the peak of the continuum emission in the lower resolution images.

The 0.13${}^{\prime\prime}$ angular resolution of these observations allow us to resolve scales of $\sim$1000\,au, the expected scale of individual star-forming cores \citep{krumholz07}. We produce a dendrogram \citep[e.g.][]{rosolowsky08} in order to describe the substructure in a systematic way. We use the {\sc astrodendro} software package to produce these dendrograms, setting a threshold of 3$\sigma$ and an increment of at least $\sigma$ between structures, where $\sigma$ is the rms sensitivity of 15 $\mu$Jy. The third parameter, minimum number of pixels required, is defined by:

\begin{equation}
\label{npix}
  N_{\rm pix}(\rm{min}) = \frac{2\pi\theta_{\rm maj}\theta_{\rm min}}{8ln(2)A_{\rm pix}} ,
\end{equation}

\noindent where $\theta_{\rm maj}$ and $\theta_{\rm min}$ are the major and minor axes of the synthesized beam and $A_{\rm pix}$ is the pixel area. For this dendrogram we used an $N_{\rm pix}$ value of 77, which corresponds to the size of approximately one synthesised beam.

\begin{figure*}
		\begin{center}
		\includegraphics[width=12cm]{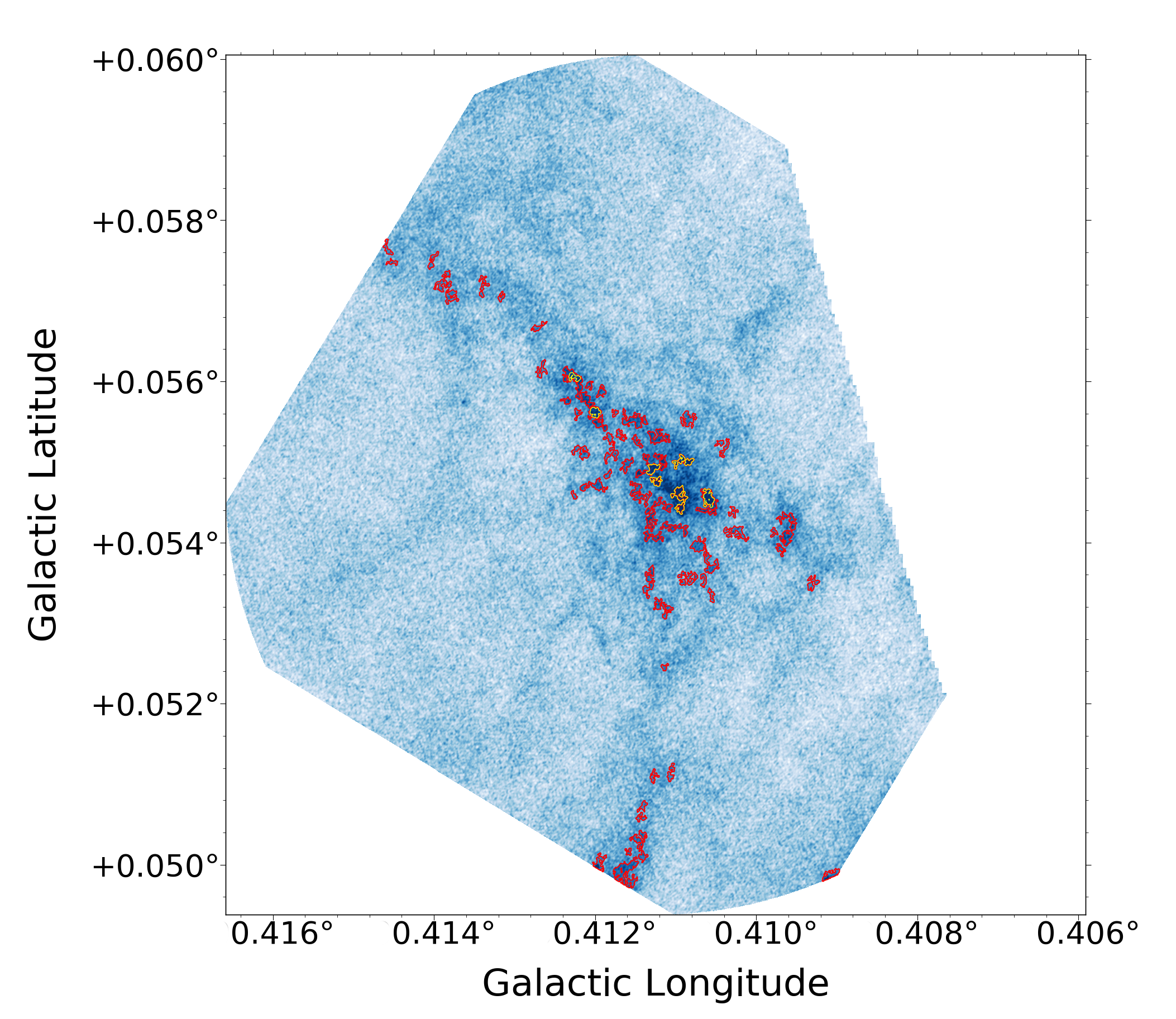}
		\caption{Locations of dendrogram leaves extracted with the {\tt ASTRODENDRO} code. The dendrogram was computed using a threshold of 3$\sigma$ ($\sigma = 16 \mu$Jy), an increment of $\sigma$ and an $N_{\rm pix}$ value of 77. Red contours show the individual compact continuum sources (`leaves') isolated by the dendrogram, of which 96 are found. Yellow contours show the nine continuum sources detected above 5$\sigma$. Note that some of the leaves extend beyond the boundary of the observed region, and so these are later excluded from analysis.}
		\label{fig:dend}
		\end{center}
\end{figure*}

Dendrograms pick out substructures as independent entities in a hierarchical manner, with the smallest possible structures being referred to as `leaves'. The dendrogram leaves are shown in Figure \ref{fig:dend}, with each red contour indicating substructure detected using the above parameters. In the context of this research, each leaf is a potential star-forming core of the scale $\sim$10${}^3$\,au (Table \ref{tab:dendfull}). Using these dendrogram parameters, we isolate 96 compact continuum sources.  We have varied the parameters of the dendrogram to see how this affects our results, namely by increasing the threshold to 5$\sigma$. In this case, we detect nine sources as opposed to 96. We repeat our analysis on just these nine sources detected over 5$\sigma$ and report our results for comparison, but for completeness we show the properties of all 96 3$\sigma$ sources.

Given the large mass of gas it is perhaps surprising that most of the structure has a column density only 3-5$\sigma$ above noise. This makes it difficult to determine whether an individual dendrogram leaf is a physically distinct object or not -- the nature of these sources is uncertain. However, it leads to one of the main conclusions of this paper, that the column density contrast in the cloud is very small. The lack of density contrast on 1000\,au scales in this cloud is reminiscent of the lack of density contrast found at 0.1 pc scales throughout the CMZ in \cite{battersby20}.

By extracting the fluxes of each leaf we calculate the mass of each source using the relation:

\begin{equation}\label{kauff}
\begin{split}
    M = 0.12M_{\odot}\Big(\textrm{e}^{1.439(\lambda/mm)^{-1}(T/10K)^{-1}}-1\Big) \\
    \indent \times \Bigg(\frac{\kappa_{\nu}}{0.01cm^2g^{-1}}\Bigg)^{-1}\Bigg(\frac{F_{\nu}}{Jy}\Bigg)\Bigg(\frac{d}{100pc}\Bigg)^2\Bigg(\frac{\lambda}{mm}\Bigg)^3,
\end{split}
\end{equation}

\noindent from \cite{kauffmann08}. Here, $M$ is the mass, $\lambda$ is the wavelength, $T$ is the dust temperature, $\kappa_{\nu}$ is the dust opacity, $F_{\nu}$ is the integrated flux and $d$ is the distance. The dust opacity towards these sources has not yet been observationally constrained, and so we estimate $\kappa_{\nu}$ using the relation:

\begin{equation}\label{kappa}
	\kappa_{\nu} = 0.04 {\textrm{cm}}^2{\textrm{g}}^{-1}\Bigg(\frac{\nu}{505 {\textrm{GHz}}}\Bigg)^{\beta} = 0.04 {\textrm{cm}}^2{\textrm{g}}^{-1}\Bigg(\frac{225{\textrm{GHz}}}{505 {\textrm{GHz}}}\Bigg)^{1.75},
\end{equation}

\noindent where $\nu$ is the frequency and the gas-to-dust ratio is assumed to be 100, which may not be the case in the CMZ \citep{longmore13,giannetti17}. The numbers used in this relation come from \cite{ossenkopf94}.

We also need to assume the dust temperature in order to estimate masses. We use a dust temperature of $\sim$20 K, based on estimates by \cite{tang20}. The uncertainties in the dust temperature and opacity mean that systematic uncertainties in the mass estimates are a factor of $\sim$2 \citep{kauffmann08}.

Additionally, we have assumed a spectral index, $\beta$, of 1.75 \citep{battersby11}. It should be noted that recent estimates by \cite{tang20} determine $\beta$ in the CMZ to be in the range 2.0 - 2.4 on scales of 10.5${}^{\prime\prime}$. Using the upper value of this range of $\beta$ = 2.4 instead of our assumed value of 1.75 increases our reported masses by a factor of $\sim$1.69. Additionally, \cite{marsh17} use Herschel data to create higher resolution maps (12${}^{\prime\prime}$) using the PPMAP procedure, reducing the average dust temperature to $\sim$17 K. If we assume this dust temperature and combine it with $\beta$ = 2.4, then our reported masses would increase by a factor of $\sim$2.10. Therefore, our mass and density estimates, using a $\beta$ value of 1.75 and dust temperature of 20 K, may be underestimates. These measures of temperature and $\beta$ are on significantly larger scales than we are probing, so any variations on smaller scales are not well constrained.

We also note that, as discussed in \cite{rosolowsky06}, sources extracted via a contour-based extraction technique often have intrinsic sensitivity and resolution biases. Interferometers systematically underestimate cloud properties, particularly the flux and by extension the mass. For example, \cite{rosolowsky06} find that the flux of Orion is underestimated by 5\% even at high sensitivity. Therefore, our mass estimates could also be underestimates by a factor of 0.05 due to this.

We calculate effective radii for each structure by calculating the radius of a circular source with an area equal to that of the structure ($R_\mathrm{eff} = \sqrt{A/\pi}$, where $A$ is the area enclosed within the dendrogram boundary). Using the calculated masses and effective radii, we then compute the volume densities of the sources, assuming a spherical geometry. 

We have assumed that all flux within each dendrogram leaf belongs to the cloud `d' continuum sources. However, the case may be that some flux in each leaf is background emission. To account for this, we have calculated background-subtracted fluxes, masses and densities for each leaf. To do this we use the minimum pixel value in each leaf as a proxy for the background emission \citep{pineda15,henshaw16b}. We then subtract this value from each pixel in the leaf and then find the total flux after this subtraction. We then use these flux values to calculate background-subtracted masses and densities. 

On average, the continuum sources with background-subtraction are $\sim$9 times less massive and dense than the non-background-subtracted continuum sources. This is likely due to the contrast within each leaf being small, and so even subtracting the minimum value integrated over the leaf is quite significant. In the rest of our analysis we use the non-background-subtracted masses and densities, but we acknowledge that these could be overestimates if any flux in the continuum sources does not belong to the source itself.

The mean mass, radius and number density of the 96 compact continuum sources are 0.67\,M${}_{\odot}$, $\sim$1.6 $\times$ 10${}^3$\,au and 7.1 $\times$ 10${}^6$\,cm${}^{-3}$ respectively, all computed from within the footprint of the dendrogram leaf. The total combined mass of the sources is 65\,M${}_{\odot}$. Repeating this analysis for the nine sources detected above 5$\sigma$ gives a mean mass, radius and number density of 0.94\,M${}_{\odot}$, $\sim$1.6 $\times$ 10${}^3$\,au and 9.7 $\times$ 10${}^6$\,cm${}^{-3}$ respectively, with a total combined mass of 8.4\,M${}_{\odot}$. A table of these values (and background-subtracted values) for each source can be found in the appendix (Table \ref{tab:dendfull}). The mass distribution is shown in Fig. \ref{fig:massdist}. We discuss the implications of these masses in Section 4.

\subsection{Search for star formation tracers within cloud `d' continuum sources}

We searched for $^{13}$CO, CH$_3$CN and SiO emission towards the cloud `d' continuum sources. $^{13}$CO is a commonly used tracer for outflows due to its high abundance and relatively low energies of the lower rotational states \citep{bally16}. CH$_3$CN is used to trace small-scale gas kinematics towards hot protostellar cores, and the relative intensities of the k-components can be used to estimate gas temperatures and column densities \citep[e.g.][]{beuther17,ilee18,maud18}. SiO is a well-known tracer of proto-stellar outflows. \citet{walker21} detected both strong CH$_3$CN emission and SiO outflows toward the young, low-mass star-forming cores in the `Brick'. However, we do not detect any CH$_3$CN emission towards the continuum sources, nor do we detect any SiO outflows. We searched for SiO outflows by manually inspecting every channel in the SiO cube and although we detect SiO emission, no distinct outflow morphology can be seen. $^{13}$CO is detected, but the emission is widespread (see the integrated intensity maps in the supplementary online material) with no distinct outflow structure. Cloud `d' also shows a lack of Class II methanol and water maser emission \citep{cotton16, rickert19, lu19b} and 70 $\mu$m emission \citep[\textit{Herschel}, HiGAL;][]{molinari10}, both tracers of star formation. Cloud `d' does not show any detections of radio continuum emission from possible H II regions \citep{immer12,lu19b}, nor does it show any 24 or 8 $\mu$m emission \citep{churchwell09}. 

In summary, even with the order of magnitude improvement in angular resolution and sensitivity provided by ALMA, cloud `d' remains unique among the dust ridge clouds in still having no signs of star formation.

\subsection{Nearest neighbour analysis of compact continuum source separations}

In order to calculate the separations of the compact continuum sources within cloud `d' and compare them to theoretically predicted values, we carry out nearest neighbour analysis using the {\tt scikit-learn} module {\tt neighbors}. We use a {\tt n\_neighbors} parameter of 2 and set the {\tt algorithm} parameter to {\tt auto}. From this analysis we find that the nearest neighbour separation between continuum sources in cloud `d' is typically of the order 10${}^3$\,au. Most continuum sources have a nearest neighbour separation of less than $\sim$7.5 $\times$ 10${}^3$\,au, with a mean of $\sim$2.6 $\times$ 10${}^3$\,au.

However, this number is not the mean separation between sources. \citet{kruijssen19b} show that the expectation value of the nearest neighbour distance is $\langle r_n \rangle = \sqrt{\pi}\lambda/4 \approx 0.443\lambda$, where $\lambda$ is the mean separation length. This expression is the integrated form of the probability distribution function of the nearest neighbour distance. Rearranging this expression for $\lambda$ means that we must multiply the mean nearest neighbour separation by a factor of $\sim$2.3 to get the mean separation length. 

This separation is also the 2D projection of the nearest neighbours, and we do not know the complete 3D separation. The implicit assumption is then that the sources all lie in the same 2D plane of the sky. In reality, the sources will also lie at different distances along the line of sight, so the separations between them may appear smaller than they are in reality due to projection effects. To correct for this, we also multiply the mean nearest neighbour separation by $\sqrt{3/2}$. This is based on the reduction of three dimensions to two, that is $(x^2+y^2+z^2)$ to $(x^2+y^2)$. 

\begin{figure}
		\begin{center}
		\includegraphics[width=\columnwidth]{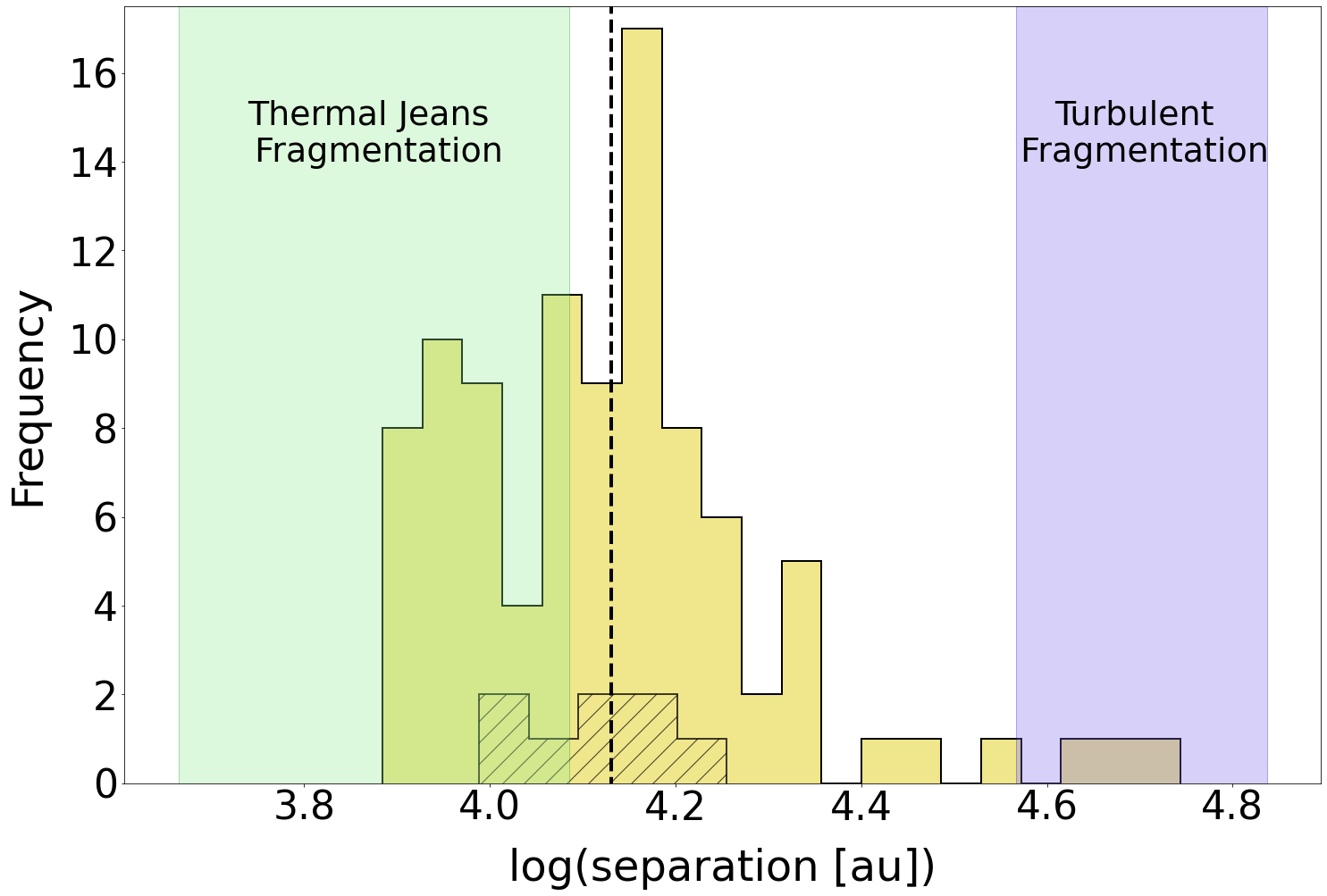}
		\caption{The separations of the 96 individual continuum sources, calculated using nearest neighbour analysis and corrected for geometric and projection effects. The black hatched histogram shows the same quantity for the nine 5$\sigma$ sources. The black dashed line shows the mean separation. The green shaded region represents the range of predicted thermal Jeans lengths and the blue shaded region represents the range of predicted turbulent Jeans lengths. The separation between compact continuum sources in cloud `d' is typically of order 10$^4$\,au, with a mean of $\sim$1.5 $\times$ 10${}^4$\,au and a standard deviation of $\sim$8.0 $\times$ 10${}^3$\,au. The separations of the continuum sources lie between thermal and turbulent fragmentation predictions.}
		\label{fig:nn}
		\end{center}
\end{figure}

Combined with the earlier factor of 2.3, this means we must multiply our mean nearest neighbour separation by a factor of $\sim$2.8. Converting each nearest neighbour separation to a separation length gives a mean separation length of $\sim$1.5 $\times$ 10${}^4$\,au. Separation lengths are shown in Figure \ref{fig:nn}. Below we compare this to the expected gas fragmentation scale.


Compared to the disc of the Galaxy, the CMZ has an elevated gas temperature \citep[60 K vs 10 K;][]{ginsburg16,immer16,krieger17} and velocity dispersion \citep[5\,kms${}^{-1}$ vs 1\,kms${}^{-1}$ at a fixed size-scale,][]{shetty12}. Therefore, we expect that the thermal and turbulent fragmentation scales in the CMZ will be different to those in the disc.

Using the effective radius (0.16 pc) and upper and lower mass estimates (239 and 69 M${}_{\odot}$, respectively) of clump `d6' from \cite{walker18}, we estimate upper and lower thermal Jeans length limits using the equation:

\begin{equation}\label{jeans}
	\lambda_\mathrm{J} \approx 0.4 {\textrm{pc}} \times \frac{c_s}{0.2 {\textrm{km s}}^{-1}} \times \Bigg(\frac{n}{10^3 {\textrm{cm}}^{-3}}\Bigg)^{-1/2},
\end{equation}

\noindent where $c_s = (k_bT/{\mu}m_H)^{1/2}$ is the sound speed and $n$ is the density of the gas. We also estimate the turbulent Jeans fragmentation length by again using Equation \ref{jeans} but this time replacing the sound speed with the velocity dispersion of clump `d6' from \cite{walker18}. They report a line width of $\sim$5 kms${}^{-1}$, which translates to a velocity dispersion of $\sim$2.1 kms${}^{-1}$.


If the gas and dust are thermally coupled, then we expect the gas temperature to be $\sim$20\,K, matching the dust temperatures reported by \cite{immer12} and \cite{walker15} of $17-23$\,K \citep{clark2013}. However, on larger scales, the gas and dust are often thermally decoupled \citep[e.g.][]{immer16,ginsburg16}. The only gas temperature constraint on similar spatial scales in this cloud is from \cite{walker18} and is an upper limit of $\sim$60\,K, based on the non-detection of higher excitation H${}_2$CO transitions towards clump `d6'. Based on this, we use upper and lower limits of 20\,K and 40\,K. We choose 40\,K as a reasonable upper limit as we do not have a firm constraint on the actual temperature, and this temperature is lower than the estimated upper limit.

Using both of these gas temperature estimates, as well as both estimates of mass, we derive thermal Jeans length estimates of $\lambda_\mathrm{J,therm}$ = 8.0${}^{+4.2}_{-3.4}$ $\times$ 10${}^{3}$\,au. The expected value of the turbulent Jeans fragmentation length is $\lambda_\mathrm{J,turb}$ = 5.3${}^{+1.6}_{-1.6}$ $\times$ 10${}^{4}$\,au. Figure \ref{fig:nn} shows the distribution of the separations, along with the mean value (black dashed line) and the ranges of the thermal and turbulent fragmentation lengths (shaded regions). The mean separation of 1.5 $\times$ 10${}^4$\,au lies between these ranges, meaning it is potentially consistent with both, although marginally more consistent with thermal Jeans fragmentation.  When making this comparison for the nine sources found above 5$\sigma$, we again find that the separations are still marginally more consistent with thermal Jeans fragmentation, with no separations being consistent with turbulent Jeans fragmentation. The mean separation, however, still lies between the two predictions, with a mean value of 1.3 $\times$ 10${}^4$\,au.

Repeating this analysis but using a value of half the thermal Jeans and turbulent Jeans fragmentation wavelengths (to represent the fact that fragments may form at half-wavelength-spaced nodes), the distribution marginally favours the turbulent fragmentation length. However, given the large uncertainty in these measurements, we conclude the data are consistent with both predictions. Therefore, it is not possible to unambiguously distinguish whether the separation distribution is more likely to be drawn from a thermal Jeans or turbulent Jeans fragmentation mechanism.

We also calculated the corresponding lower and upper limits of the thermal Jeans mass and turbulent Jeans mass using the equation:

\begin{equation}\label{jeansmass}
	M_\mathrm{J} \approx 2 {\textrm{M}}_{\odot} \times \Bigg(\frac{c_s}{0.2 {\textrm{km s}}^{-1}}\Bigg)^3 \times \Bigg(\frac{n}{10^3 {\textrm{cm}}^{-3}}\Bigg)^{-1/2}.
\end{equation}

The estimated thermal Jeans mass is $M_\mathrm{J,therm}$ = 0.54${}^{+0.70}_{-0.35}$ $M_{\odot}$. The estimated turbulent Jeans mass is $M_\mathrm{J,turb}$ = 144${}^{+43}_{-43}$ $M_{\odot}$. Therefore, the masses of the sources (see Table \ref{tab:dendfull}) are roughly consistent with the predicted thermal Jeans mass limits of the clump and not with the turbulent Jeans mass predictions. When making this comparison for the nine sources found above 5$\sigma$, we find the same trend - the majority of sources are consistent with thermal Jeans mass estimates, with a few lying at masses greater than this.

In summary, the consistency of the source masses with thermal Jeans mass estimates, along with the slight tendency of the separations towards thermal Jeans fragmentation, it seems as though thermal Jeans fragmentation is a better descriptor of the source structure in this cloud. This is consistent with recent results from \citet{lu20} and \citet{walker21}.


\subsection{Virial analysis of cloud `d' continuum sources}
\label{sec:vir}

We produce spectra towards each of the continuum sources (see supplementary materials) in SiO and three different H${}_2$CO transitions. To produce these spectra, we extract an averaged spectrum for each leaf in the dendrogram. We then use the {\tt pyspeckit} Python package to fit each these spectra with a Gaussian profile in order to obtain velocity dispersions. We set the {\tt guesses} parameter as: the amplitude is equal to the maximum amplitude within the spectrum, the velocity is equal to the velocity at the point of highest amplitude, and the width is equal to full width half maximum of the line. We note that only H${}_2$CO effectively traces the mass (see supplementary online material), and so we only use the lowest energy formaldehyde transition in later analysis. 

We find that only 13 of the continuum sources have clearly detected spectra, and we calculate the virial parameters of them using the equation:

\begin{equation}\label{virial}
\alpha = \frac{5R\sigma^2}{GM},
\end{equation}

\noindent from \cite{bertoldi92}, where $R$ is the effective radius of each continuum source, $\sigma$ is the velocity dispersion and $M$ is the calculated upper mass limit. A value of $\alpha \lesssim$ 2 indicates that a body is gravitationally bound. 

For the majority of continuum sources, we could not measure an appropriate velocity dispersion, and the ones that we could measure have large measurement uncertainties. For continuum sources where we could measure reliable velocity dispersions, we calculate $\alpha$ values in the range 5${}^{+13}_{-4.94}$ - 45${}^{+55}_{-33}$ (see Table \ref{tab:virial} in the appendix for a full list of these calculated values). All of the calculated values are greater than 2. However, their uncertainties mean that some may have a virial parameter of less than 2. Therefore, assuming that these values are representative of the whole sample of continuum sources, it may be the case that some of the continuum sources in cloud `d' are gravitationally bound, but uncertainties on the velocity dispersions of the continuum sources mean that in practice the virial state of the sources is essentially unconstrained.

If we use only the nine 5$\sigma$ sources detected, then only one of the sources has reliable H${}_2$CO emission. Therefore, we continue virial analysis with the full 96 source sample, as this still leaves the virial state of the sources unconstrained and ultimately does not change our conclusions.

This analysis of the virial parameter faces a potential issue, in that H${}_2$CO emission is extended. When we extract the spectrum towards continuum sources, it is important to distinguish the relative contribution from the compact continuum source and larger scales. Including emission from gas not associated with the source will affect the inferred gravitational boundedness. However, we do know that the formaldehyde and dust emission have good spatial correspondence, so we assume that this effect is minimal. Additionally, while H${}_2$CO is extended, it is used as a reliable tracer of dense gas kinematics in the CMZ \citep[e.g.][]{walker18, lu21}, and so we consider it a reliable tracer of velocities.

\begin{figure}
		\begin{center}
		\includegraphics[width=\columnwidth]{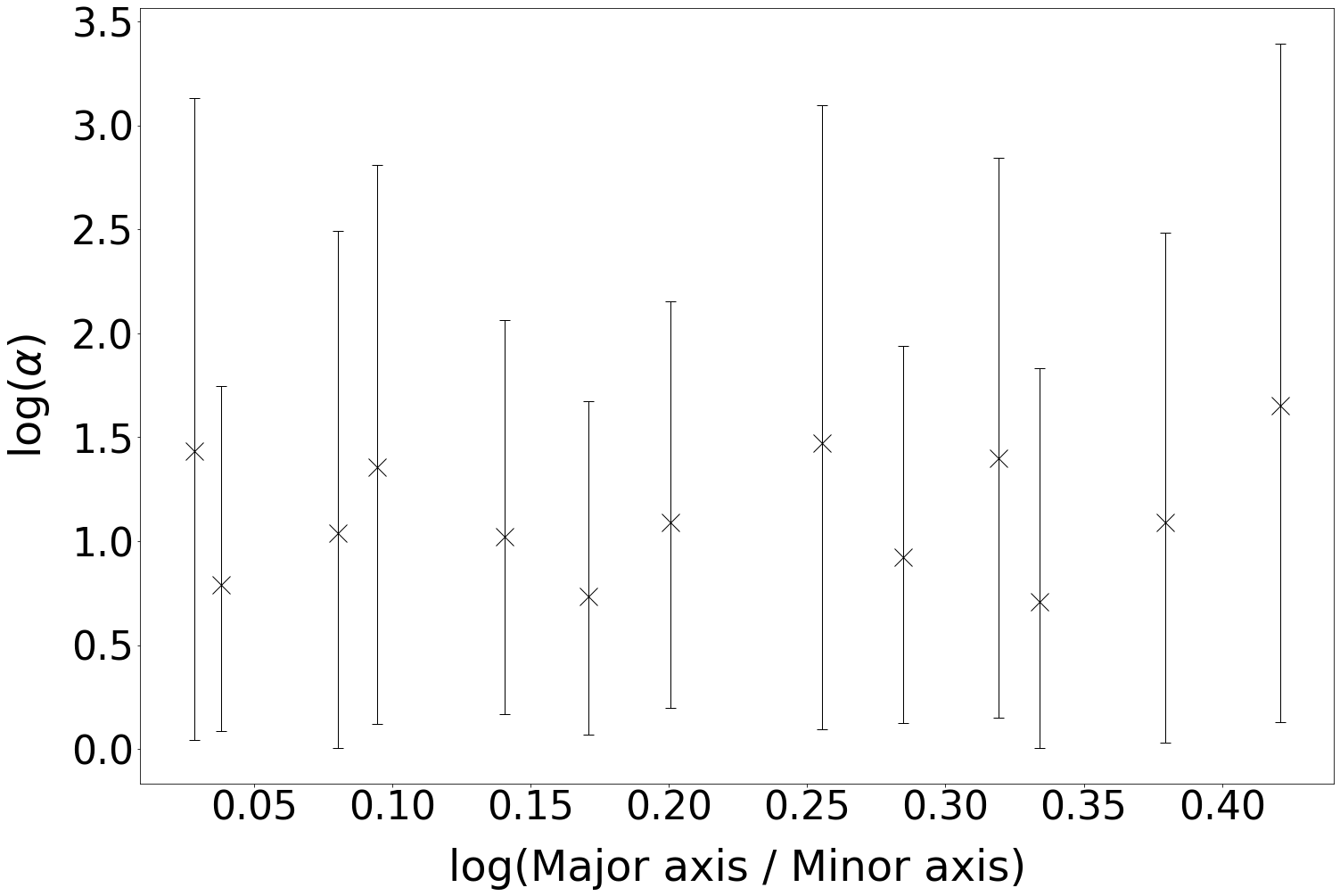}
		\caption{Comparison of the virial parameters of the continuum sources for which we could measure a reliable velocity dispersion and their aspect ratios. One might expect continuum sources with a lower virial parameter (i.e. more likely to be gravitationally bound) to be more circular, but there is no correlation between these parameters.}
		\label{fig:asprat}
		\end{center}
\end{figure}

To investigate the virial state further, we find the elongation of each continuum source by calculating the ratio of the major axis to the minor axis. We expect that unbound structures will have a higher degree of elongation than bound structures. This is because gravitationally bound structures would appear roughly circular due to being roughly spherical, whereas unbound structures are less likely to appear circular. We find ratios in the range of 1 - 4.5, with a mean value of $\sim$1.92. We also plot the virial parameter against the aspect ratio (Figure \ref{fig:asprat}) to see if continuum sources with a lower virial parameter correspond to a smaller aspect ratio. However, there is no clear correlation. 

In Section 4.3, we discuss the possibility of these continuum sources also being in pressure-bounded equilibrium.

\section{Discussion}

Dust ridge cloud `d' is one of the most massive (10$^{4-5}$~M$_{\odot}$) and compact (R $\sim$ 3~pc) molecular clouds known to exist in the Galaxy in which there are no known signs of local or widespread star formation. Our ALMA observations towards the highest-density region of the cloud confirm that there is no evidence of active star formation in the form of outflows or gas tracers down to protostellar scales (1000\,au).

Our analysis of the continuum emission reveals an overall lack of compact substructure in the cloud. We identify a population of 96 low-mass continuum sources, for which the gas structure is more likely to be set by thermal  fragmentation rather than turbulent fragmentation (Figure \ref{fig:nn}). In $\S$~\ref{sec:vir} we found that the majority of these continuum sources are unlikely to be gravitationally bound ``cores". Moving forward in the paper, we will therefore continue to refer to them as ``sources".

Nearest-neighbour analysis has shown that the sources have separations consistent with both thermal Jeans and turbulent Jeans fragmentation, with a tendency towards thermal. Additionally, the masses of the sources are  consistent with the predicted thermal Jeans mass of the cloud. Several recent studies \citep{lu19b,lu20,walker21} have found that separations are consistent with thermal Jeans fragmentation in other CMZ clouds, including the `Brick'. 

This suggests that, while turbulence drives gas properties on large-scales in the CMZ, smaller scales may be less sensitive to this. On protostellar scales, star formation in the CMZ may proceed in a similar way to star formation regions in the local neighbourhood, albeit with a higher critical density threshold to overcome before star formation can begin \citep{walker18, barnes19}.

\begin{figure}
		\begin{center}
		\includegraphics[width=\columnwidth]{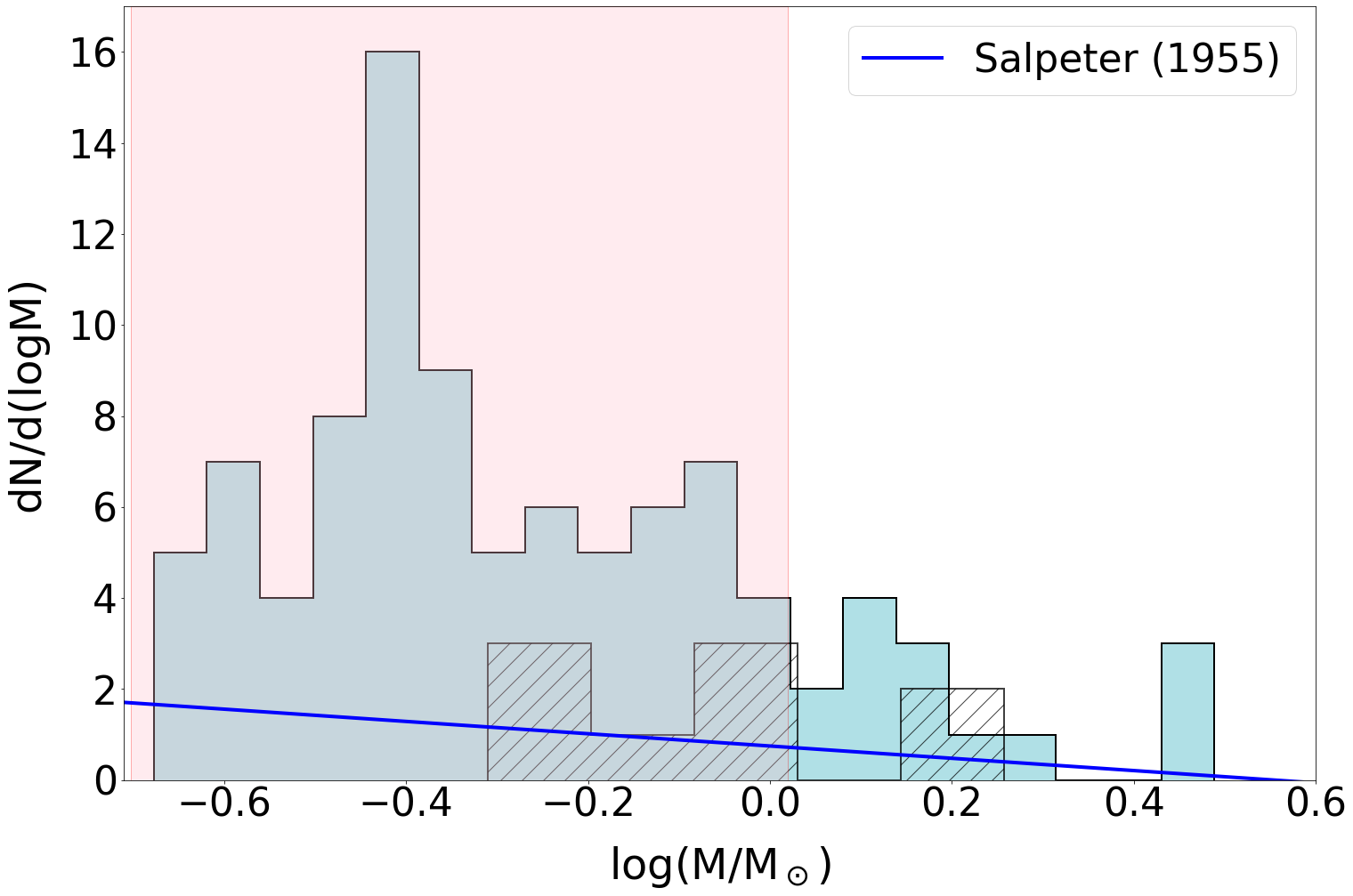}
		\caption{Mass distribution of the 96 individual sources detected by our dendrogram analysis. The black hatched histogram shows the same quantity for the nine 5$\sigma$ sources. The \citet{salpeter55} IMF is overplotted (blue line). The pink shaded region shows the range of predicted thermal Jeans masses for the clump, for which most masses are consistent.}
		\label{fig:massdist}
		\end{center}
\end{figure}

Figure~\ref{fig:massdist} shows a mass distribution of the sources. Compared to a standard core mass function, the distribution is bottom-heavy and has no high-mass progenitors -- the most massive source is $\sim3$\,M$_{\odot}$, the mean mass of $\sim$0.7~M$_{\odot}$, and most sources being $<1$\,M${}_{\odot}$. If these sources are the precursors to stars, in order for the resulting stellar distribution to conform to a normal IMF, the compact continuum sources must gain many times their current gas mass from the surrounding environment. Calculating the mass distribution for the nine sources detected above 5$\sigma$, the majority of sources are still $<1$\,M${}_{\odot}$, meaning they would still have to accrete a large amount of material from the surrounding environment. However, in both cases, not much can be said about the slope of this mass distribution as there are not enough sources to build a statistically meaningful mass function.


Coupled with the lack of molecular line emission tracing hot cores or outflows, our results paint a coherent picture of a massive molecular cloud with no signs of star formation. In the following, we compare our results with complementary data to investigate the ultimate fate of this extreme molecular cloud.

\subsection{Do the properties of sources vary with environment?}


On global ($1 - 100+$\,pc) scales in the CMZ, gas conditions are known to be extreme compared to the Solar neighbourhood. Densities \citep[$\sim$10${}^{3-4}$ cm${}^{-3}$,][]{guesten83, mills18}, gas temperatures \citep[typical gas temperatures are $50-100$\,K, but can reach as high as $400-600$\,K,][]{mills13, ginsburg16}, pressures \citep[$P/k_B \sim$10${}^{7-9}$\,K\,cm${}^{-3}$,][]{longmore14,rathborne14} and line widths \citep[$\sim10-20$\,km\,s${}^{-1}$,][]{henshaw16} are between several factors to several orders of magnitude greater than those found in the disc. It is therefore plausible that the star formation process may occur differently in such an environment. To investigate this on protostellar (1000\,au) scales, we compare the properties of the sources detected in cloud `d' and compare them with cores both in the CMZ and the Galactic disc.

\begin{figure*}
		\begin{center}
		\includegraphics[width=18cm]{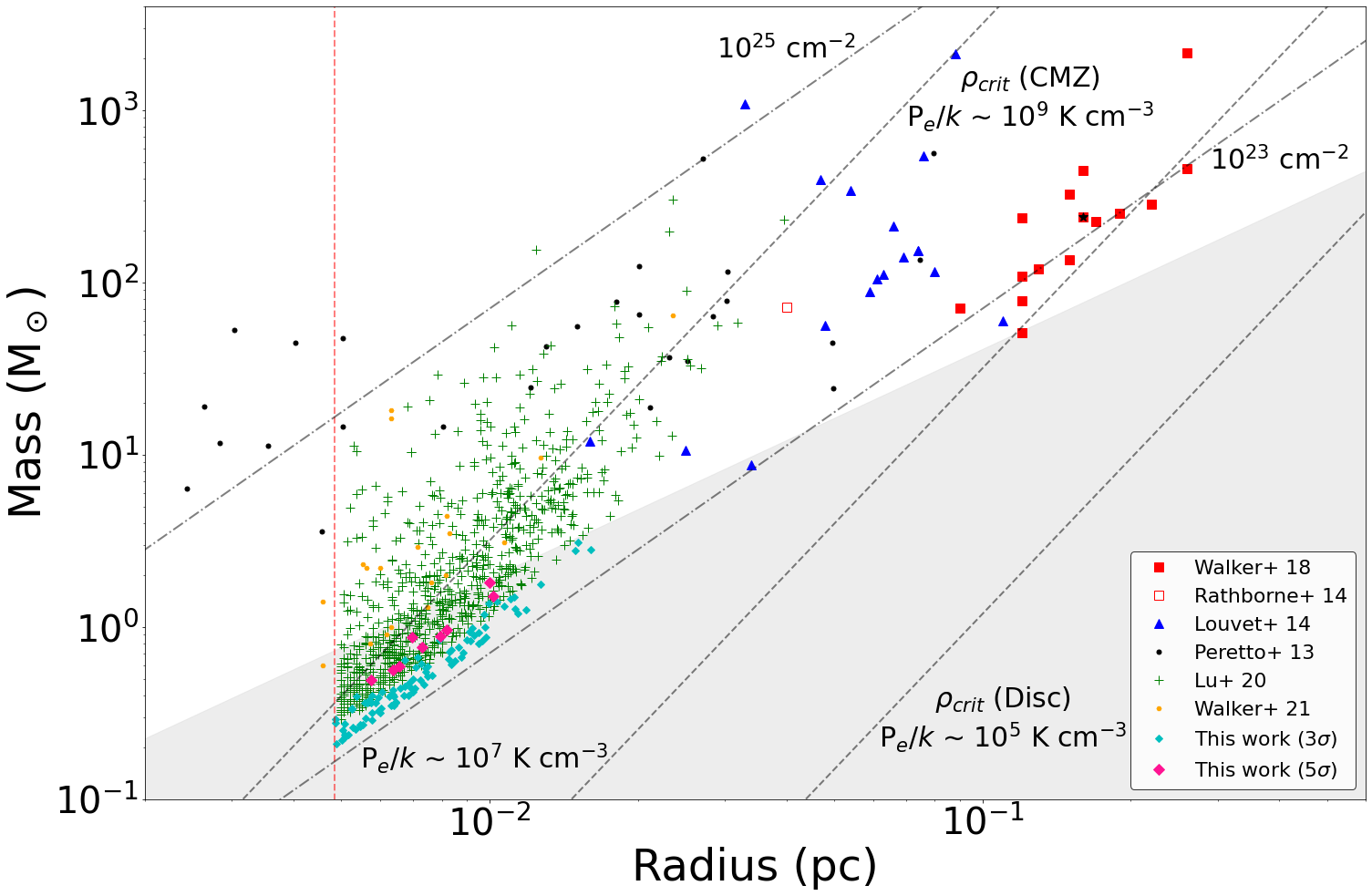}
		\caption{Mass-radius plot for all of the compact continuum sources reported in our ALMA sample. The solid cyan diamonds correspond to masses of the 96 sources detected above 3$\sigma$, estimated assuming a dust temperature of 20K. Solid pink diamonds show the same quantity for the nine sources detected above 5$\sigma$. Solid red squares correspond to SMA clump upper mass limits reported in \citet{walker18}. The red point with a black star marker indicates the clump from \citet{walker18}'s sample that we have observed in this work. Black points correspond to high-mass protostellar cores in the Galactic disc taken from \citet{peretto13} and blue triangles to those from Louvet et al. (2014). The open red square corresponds to the star-forming core in cloud `a' (aka the `Brick') as seen with ALMA observations \citep{rathborne14}. Green crosses represent cores within four CMZ clouds as seen with ALMA observations \citep{lu20} and yellow points correspond to cores found within cloud `a' by \citet{walker21}. Dash/dot lines show constant column density. Dashed lines show the predicted critical volume density thresholds for both the CMZ and the Galactic disc, assuming pressures of $P/k$ = 10${}^9$ and 10${}^5$\,K\,cm${}^{-3}$, respectively, with an intermediate threshold for a pressure of $P/k$ = 10$^7$\,K\,cm$^{-3}$. The red dashed line corresponds to our resolution of $\sim$1000\,au. The area above the grey shaded region corresponds to the  portion of the mass-radius plane above the empirical high-mass star formation threshold proposed by \citet{kauffmann10}.}
		\label{fig:mrplot}
		\end{center}
\end{figure*}

We take Figure 3 from \cite{walker18}, which shows the mass-radius plot for a sample of CMZ and disc cores from the literature, and add our 96 sources. Also plotted are \cite{walker18}'s 15 SMA clumps and the cloud `a' core (aka the `maser core' in the `Brick') mass and radius from \cite{rathborne14}. The plot also includes a sample of high-mass protostellar cores found in the Galactic disc, taken from \cite{peretto13}, as well as high-mass protostellar cores in the W43-MM1 ridge, a likely precursor to a `starburst cluster', taken from \cite{louvet14}. The masses of these cores have been scaled to make them consistent with the spectral index of $\beta$ = 1.75 that we have used for mass estimates in our analysis. We have also plotted masses and radii for multiple CMZ cores from \cite{lu20} and the masses and radii of eighteen cores in cloud `a' \citep[aka the `maser core' in the `Brick'/G0.253$+$0.016,][]{walker21}. The masses and radii of the cores in these samples were both calculated in the same way as the ALMA sources, assuming the same dust temperature of 20 K, and so direct comparison is possible. The resulting plot is shown in Figure~\ref{fig:mrplot}.

We find that the sources detected in cloud `d' are consistent with the mass-radius relationship of cores detected in star-forming clouds in both the CMZ and the disc, though the cloud `d' sources are on the lower end of the mass distribution at a given size scale. The area above the grey shaded region of Fig. \ref{fig:mrplot} corresponds to cores which lie above the empirical massive star formation threshold proposed by \cite{kauffmann10}, which they determine to be $M(R) \gtrsim$ 870 M${}_{\odot} \times (R$/pc)${}^{1.33}$. We find that all of the cloud `d' sources are below this limit, and therefore should not be forming high-mass stars. Assuming that the sources are not transient and continue to accrete mass, they may eventually exceed this threshold and begin forming stars. However, as \cite{kauffmann10}'s threshold was derived for clouds in the Solar neighbourhood, it is unclear whether this should hold in the CMZ. Indeed, \cite{walker18} find that all of their CMZ clumps are on or above this limit, but only two show signs of star formation. At the scale of our ALMA data, it is difficult to conclude whether or not there is an environmental dependence on star formation, as none of our sources are above the threshold.

\subsection{Are the sources in hydrostatic equilibrium?}

Although the virial ratios calculated in Section 3.4 suggest the sources may not be self-gravitating, previous observations have suggested dense gas sources in the CMZ may be in hydrostatic equilibrium but confined by the high ambient pressure in the CMZ. We follow the analysis of both \cite{field11} and \cite{walker18} to investigate this possibility in cloud `d'. \cite{field11} study Galactic disc clouds from the Galactic Ring Survey \citep[GRS,][]{jackson06} as self-gravitating isothermal spherical clouds, subject to a uniform external pressure $P_e$, in the context of the virial theorem. Based on the GRS analysis of \cite{heyer09}, the disc clouds are not bound when considering simple virial equilibrium. \cite{field11} find that external pressures of $P_e/k\sim$10${}^{4-6}$ K cm${}^{-3}$ acting upon the clouds are needed for them to be in pressure-bounded equilibrium.

\cite{walker18} expanded on the \cite{field11} analysis by comparing their SMA observations of dust ridge clouds and clumps with the disc clouds. They conclude that, if these dust ridge clouds and clumps are in pressure equilibrium, then the external pressures in the CMZ would have to be of order $P_e/k\sim$10${}^{8}$ K cm${}^{-3}$, 2-3 orders of magnitude greater than necessary for the clouds in the Galactic disc. This is consistent with the measured ambient pressure in the CMZ of $P_e/k\sim$10${}^{7-9}$ K cm${}^{-3}$ \citep{longmore14, rathborne14,kruijssen14}.

\begin{figure*}
		\begin{center}
		\includegraphics[width=18cm]{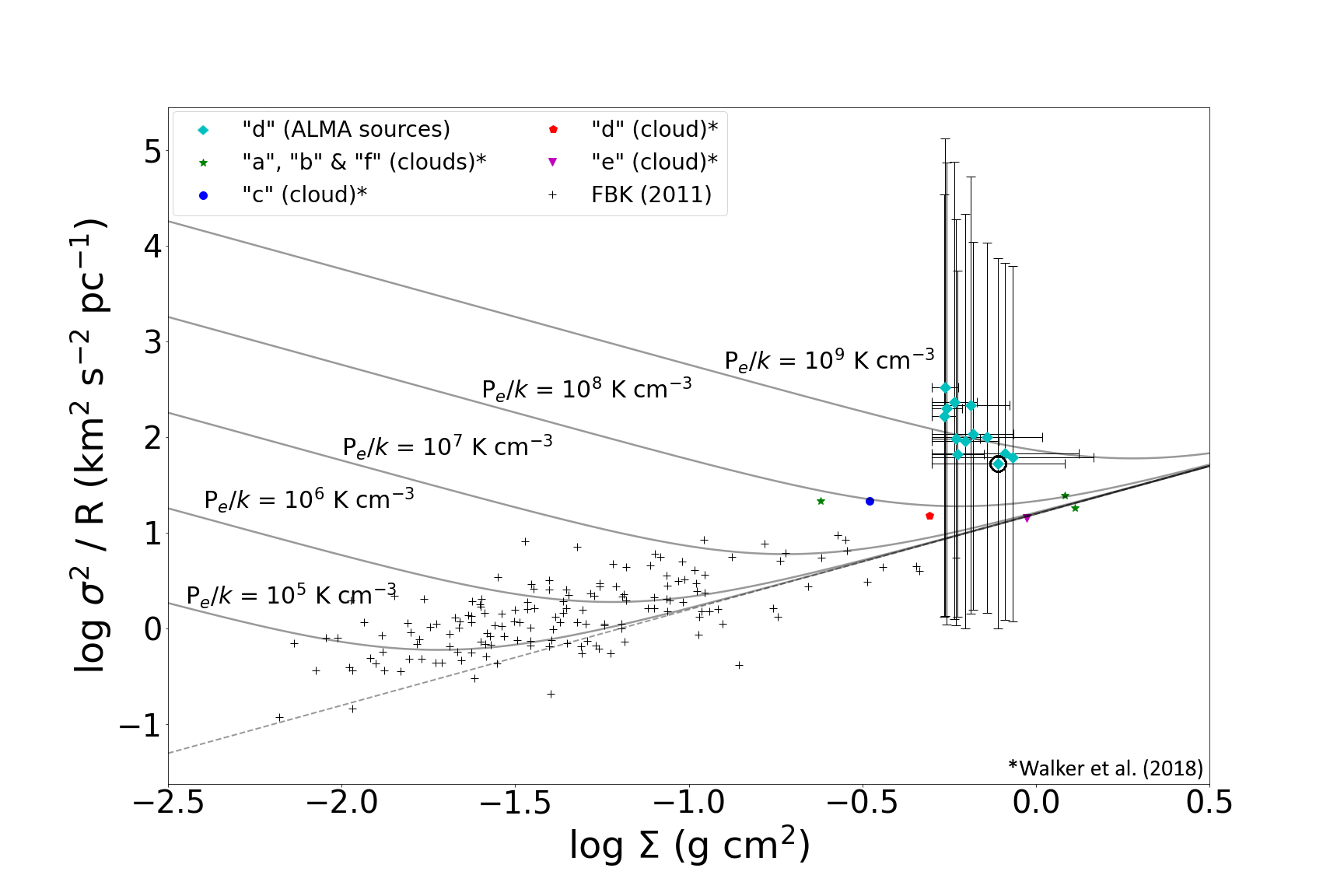}
		\caption{Comparison of the dust ridge sources with dust ridge clouds and GRS clouds. Black crosses show disc clouds as reported in \citet{field11} \citep[original data from][]{heyer09}. Diamond shaped markers indicate the cloud `d' sources. The data point circled in black is the only source that was detected above 5$\sigma$ that also had reliable H${}_2$CO emission. All other solid markers indicate dust ridge clouds, as reported in \citet{walker18}. Curved black lines are those of constant external pressure, while the dashed line is for $P_e$ = 0.}
		\label{fig:h2copress}
		\end{center}
\end{figure*}

We further expand on this by comparing our ALMA compact dust continuum sources to both the GRS data and the SMA dust ridge clouds in the context of pressure equilibrium \citep{walker18}. Figure~\ref{fig:h2copress} shows a replica of Figure 3 in \cite{field11}, along with the SMA dust ridge cloud data in \cite{walker18}. We have additionally overplotted our cloud `d' sources. Note that only a subset of the 96 sources detected are displayed, as a measure of line-width is required, which was not possible towards all of our ALMA sources due to a lack of significant molecular line emission. The dashed black line represents  virial equilibrium, where no external pressure is present. Each of the curved black lines represent pressure-bounded virial equilibrium for several different external pressures. These lines are described by a reformulated version of the pressure-bounded virial equation:

\begin{equation}\label{pve}
	V_0^2 = \frac{\sigma^2}{R} = \Bigg(\pi\Gamma G\Sigma + \frac{4P_e}{\Sigma}\Bigg),
\end{equation}

\noindent where $V_0$ is the size-line-width scaling coefficient, $\sigma$ is the velocity dispersion, $R$ is the radius, $\Gamma$ is related to the density structure \citep{elmegreen89, field11}, $\Sigma$ is the mass surface density and $P_e$ is the external pressure. We assume $\Gamma$ = 0.73, corresponding to a centrally concentrated density structure. This is likely valid for cores of this scale, but not for the clouds on larger scales, as they display relatively flat surface density profiles \citep{walker15, walker16}.

From Fig. \ref{fig:h2copress} we can see that the pressures required to confine sources on the scale of these sources are $\sim$10${}^9$\,K\,cm${}^{-3}$. This is an order of magnitude larger than the external pressures determined for the large-scale clouds by \cite{walker18}. This suggests that there is either an additional confining pressure on source scales that is undetected in cloud scale observations, or that the sources are over-pressured with respect to the surrounding gas and are therefore transient.

The densities of these sources are of the range 10$^{6-7}$\,cm$^{-3}$, 2 to 3 orders of magnitude greater than the volume density threshold proposed by \cite{lada10}. Despite this, and the high external pressures the sources are subject to in the CMZ, the sources show no signs of star formation. We conclude that this is further evidence for star formation being inhibited in the CMZ due to the critical volume density threshold for star formation being driven up by the high turbulent energy density in this environment \citep{kruijssen14,rathborne14}.


\subsection{Evidence for Convergent Gas Flow}
\label{sub:cgf}

The lack of star formation in cloud `d' indicates one of two scenarios -- either cloud `d' is in the very early stages of cluster formation, or it will never form a cluster at all. Given the high gas density \citep[$n_{\rm H_2}$ = 0.7 $\times$ 10$^{5}$ cm${}^{-3}$, $t_{\rm ff}$ = 11.4 $\times$ 10$^{4}$ yr;][]{barnes19}, the fact the cloud and the dense clumps (at the scale $R_\textrm{eff} \sim$0.1 pc) within it are gravitationally bound \citep{longmore13b, barnes19}, and the short free-fall times at all scales \citep{barnes19}, the latter of these scenarios would pose a serious challenge to star formation theories. 

To distinguish between these scenarios, we now try to better understand the likely fate of cloud `d'.


\subsubsection{Gas flows at the cloud scale}

We use MALT90 data \citep{foster11,foster13,jackson13}, specifically the HNCO (4$_{0,4}$ - 3$_{0,3}$) and SiO (2-1) lines, to investigate the gas kinematics in cloud `d' at a larger scale. HNCO is commonly used as a reliable dense gas tracer in the CMZ \citep{henshaw16,henshaw19} and SiO is a well-known tracer of shocked gas. The larger scale MALT90 data has an angular resolution of 40$^{\prime\prime}$ -- two orders of magnitude larger than that of our ALMA data -- providing the pc-scale gas motion in and around cloud `d'.

\begin{figure*}
		\begin{center}
		\includegraphics[width=18cm]{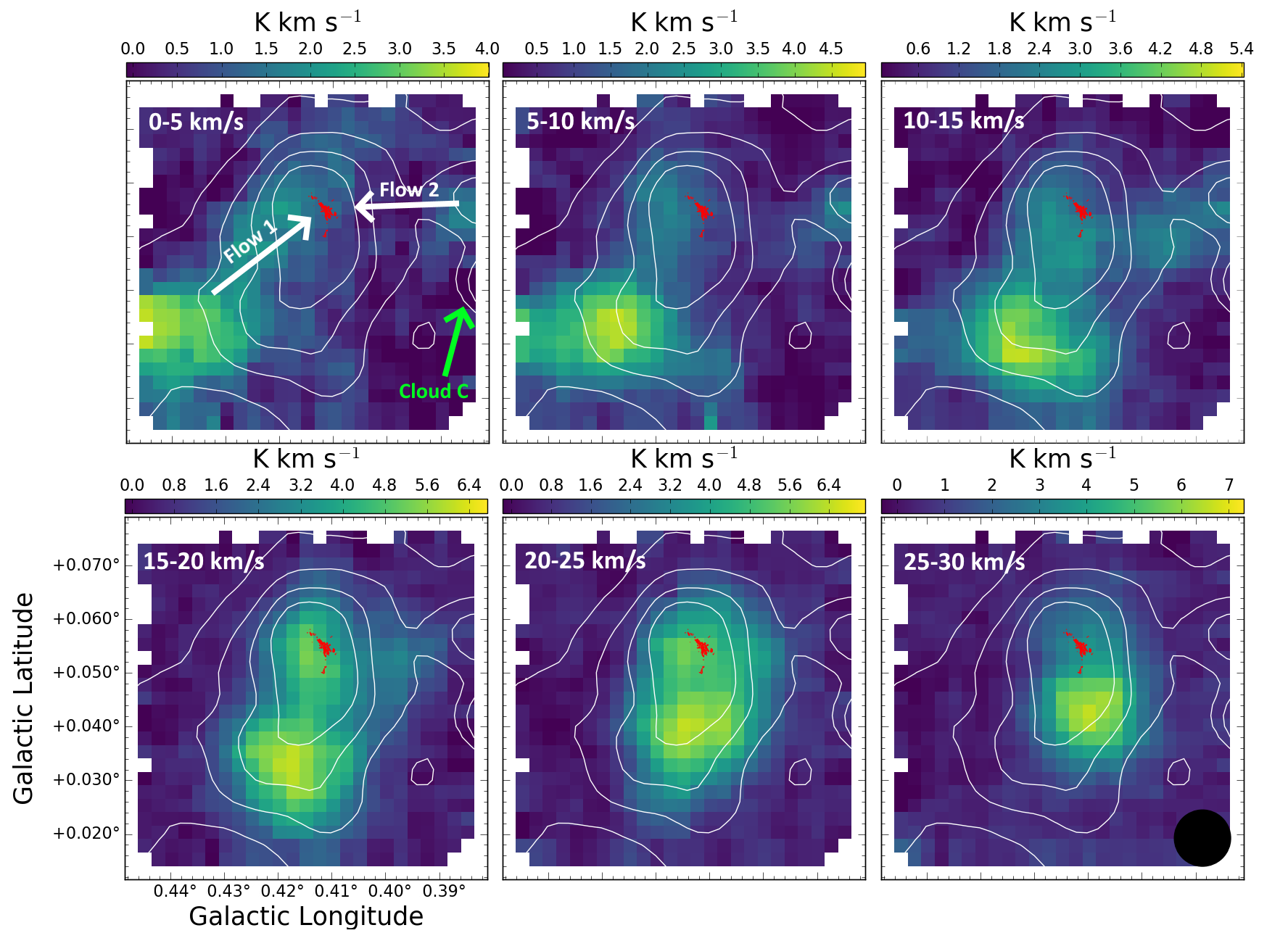}
		\caption{Mopra 22-m telescope data of HNCO emission towards cloud `d' as part of the MALT90 Survey. The angular resolution of this data is 40$^{\prime\prime}$. Red contours show the 0.13$^{\prime\prime}$ ALMA data presented in this paper. Black contours show BGPS data. The black circle in the bottom right plot represents the primary beam of the Mopra data. The data shows channel maps of HNCO in the velocity range $0 - 30$\,kms$^{-1}$. Part of cloud `c' is visible to the right of each map. At low velocities the HNCO emission is found to the right and bottom left of the continuum emission peak (outlined by the ALMA contours). As the velocity increases, the emission from the right and bottom left both steadily move towards the continuum peak. The convergence of this velocity gradient at the location of the continuum peak is the same kinematic signature found by \citet{henshaw16} in the gas upstream from the dust ridge. We interpret this kinematic structure as the convergence of pc-scale gas flows at the continuum peak of cloud `d'.}
		\label{fig:hncom90}
		\end{center}
\end{figure*}

\begin{figure*}
		\begin{center}
		\includegraphics[width=18cm]{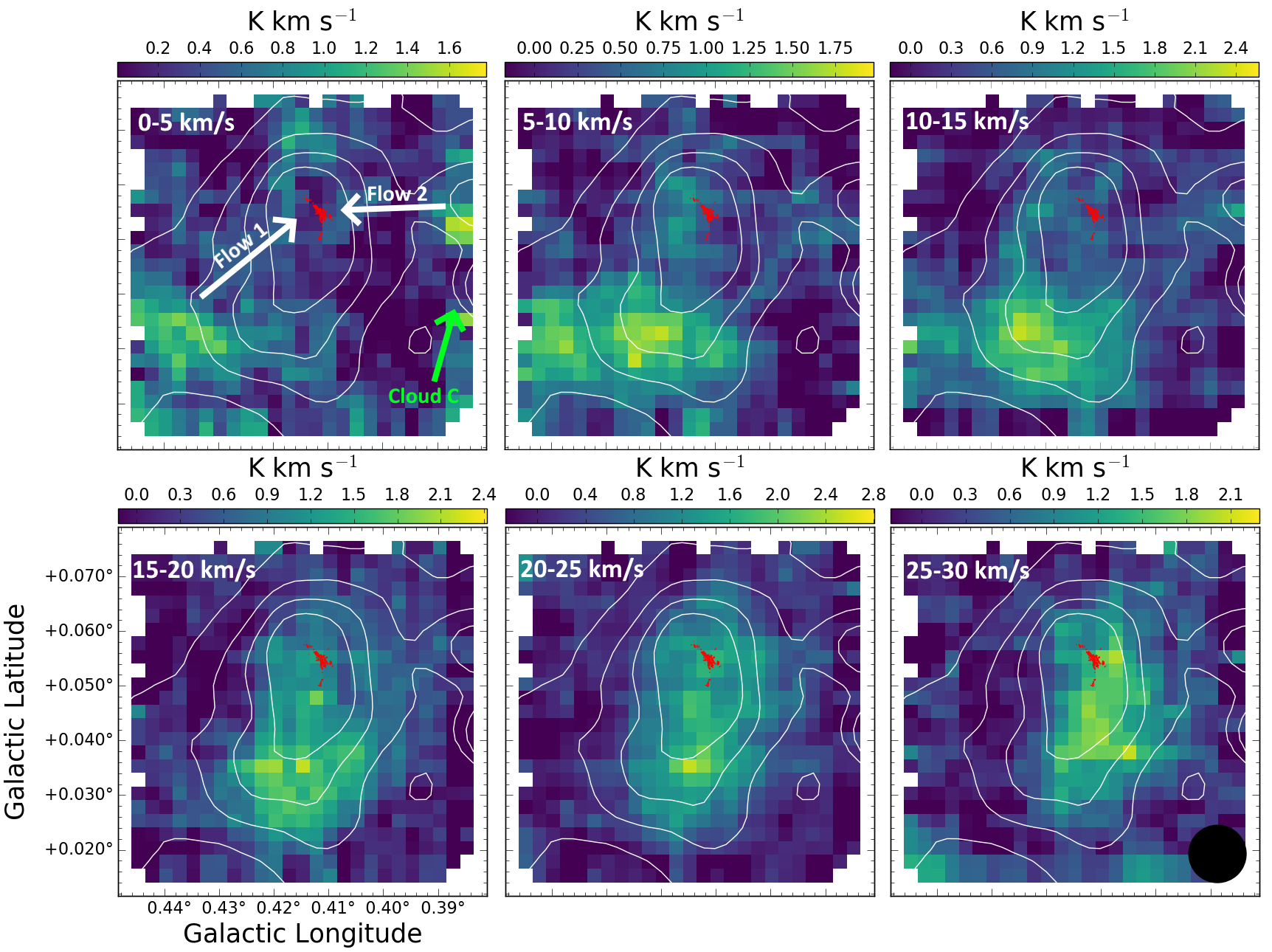}
		\caption{Mopra 22-m telescope SiO(2-1) data of cloud `d' as part of the MALT90 Survey. The angular resolution of this data is 40${}^{\prime\prime}$. Red contours show our 0.13${}^{\prime\prime}$ ALMA data and black contours show BGPS data. The black circle in the bottom right plot represents the synthesised beam. The data shows channel maps of SiO in the velocity range 0 - 30 kms${}^{-1}$. Part of cloud `c' is visible to the right of each map. A clear velocity gradient can be seen, with a point of convergence between clouds `d' and `c' at the point of our ALMA data.}
		\label{fig:siom90}
		\end{center}
\end{figure*}

Figures \ref{fig:hncom90} and \ref{fig:siom90} show the gas motions of HNCO and SiO in cloud `d' respectively. To make these plots, we map integrated intensity in spacings of 5 kms${}^{-1}$ in the range $0 - 30$\,kms${}^{-1}$. Another dust ridge cloud (cloud `c') can also be seen at the far right of each map. Red contours show the 0.13$^{\prime\prime}$ ALMA 1.3~mm dust continuum data for scale. Black contours show the BOLOCAM Galactic Plane Survey data for clouds `d' and `c'. 

The following interpretation of the three-dimensional motion assumes that the velocity along the line of sight is comparable to the velocity in the plane of the sky. At low velocities of $0-10$\,km${}^{-1}$, the HNCO and SiO emission is found to the right and bottom left of the continuum emission peak (outlined by the ALMA continuum contours). As the velocity increases, the emission from the right and bottom left both steadily move towards the continuum peak, converging at this location at a velocity of $\sim20-25$\,kms$^{-1}$.


In addition to these large-scale Mopra channel maps, we produce channel maps of the ALMA data. Figures \ref{fig:h2cochan} and \ref{fig:siochan} show the H${}_2$CO (3${}_{0, 3}$-2${}_{0, 2}$) and SiO (5-4) emission, respectively, at velocities in the range $5 - 30$\,kms${}^{-1}$.  The spatial morphology of the H$_{2}$CO emission is dominated by two partially filled circular structures which intersect at the location of the dust continuum emission at a velocity of $15 - 20$\,kms$^{-1}$. The SiO emission also shows a similar morphology, although less of both circles are filled, possibly due to the lower signal-to-noise of the detection. A key difference between the SiO and H$_{2}$CO morphologies is that the emission from the two overlapping H$_{2}$CO circles coincides exactly with the dust emission, while the location of the overlapping circles in the SiO emission is offset to the right of the dust emission. Indeed, the SiO emission appears to be ``wrapping around" the right-hand edge of the dust continuum emission. This spatial offset is consistently equivalent to $\sim$7 beam widths, so must be real and not an observational artefact. No signs of SiO bi-polar outflows indicative of ongoing star formation are detected.


We now try to interpret this information in a self-consistent way. The convergence of the large-scale HNCO and SiO velocity gradients at the location of the continuum peak is similar to the kinematic signature found in similar Mopra data by \citet{henshaw16c} in the gas upstream from the dust ridge. They showed that this velocity structure can arise from the convergence of large scale gas flows due to global gravitational collapse. A natural interpretation of the HNCO and SiO kinematic structure in the Mopra data towards cloud `d' is therefore that it is showing convergence of pc-scale gas flows at the continuum peak.

In this scenario, the curved ALMA dust continuum structure (at clump `d6') is the most likely convergent point of the flows, with mass converging from across the entirety of cloud `d'. Given that the relative motion of the gas flows is highly supersonic, they should produce strong shocks at the intersection point. The fact that the SiO emission curves around the right-hand edge of the dust continuum emission provides strong evidence that there is a shock front at this location at V$_{\rm LSR} = 15 - 20$\,kms$^{-1}$.

As the ambient gas in the flows is already at high density, it should cool quickly after the shock front has passed, making this an isothermal shock. The post-shock density will therefore be enhanced by a factor $\cal{M}$$^2$, where $\cal{M}$ is the Mach number of the shock \citep{padoan11}. However, it is possible that C-shocks are present, and so the compression is not necessarily given by $\cal{M}$$^2$. The exact density enhancement will depend on the details of the shock. It is plausible that the gas traced by the ALMA dust continuum emission reached its high density through this process.

Here, the two unfilled circles in the ALMA H$_{2}$CO channel maps represent the excitation- and density-enhanced gas along the intersection point of the two flows which converge at $15 - 20$\,kms$^{-1}$ at the location of the dust continuum emission. The offset of the SiO emission to the right-hand side of the dust continuum emission shows the current location of the shock front. The curvature and central peak of the dust emission and the gas is strikingly similar to the bow shock morphology seen ubiquitously towards shocked regions.

\begin{figure*}
		\begin{center}
		\includegraphics[width=18cm]{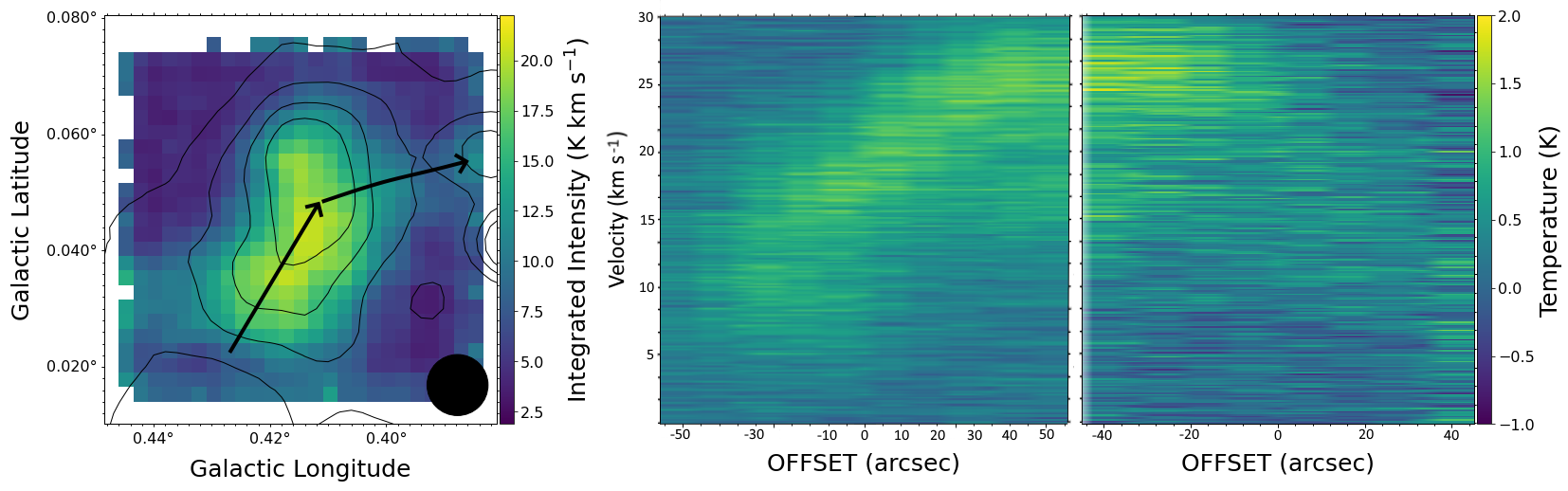}
		\caption{Left: Integrated intensity map of the Mopra HNCO line. Black arrows show the slices that the pv diagrams were computed across. Middle and right: Position-velocity (pv) diagrams of the two flows within the Mopra HNCO line.}
		\label{fig:pv}
		\end{center}
\end{figure*}

\begin{figure*}
		\begin{center}
		\includegraphics[width=18cm]{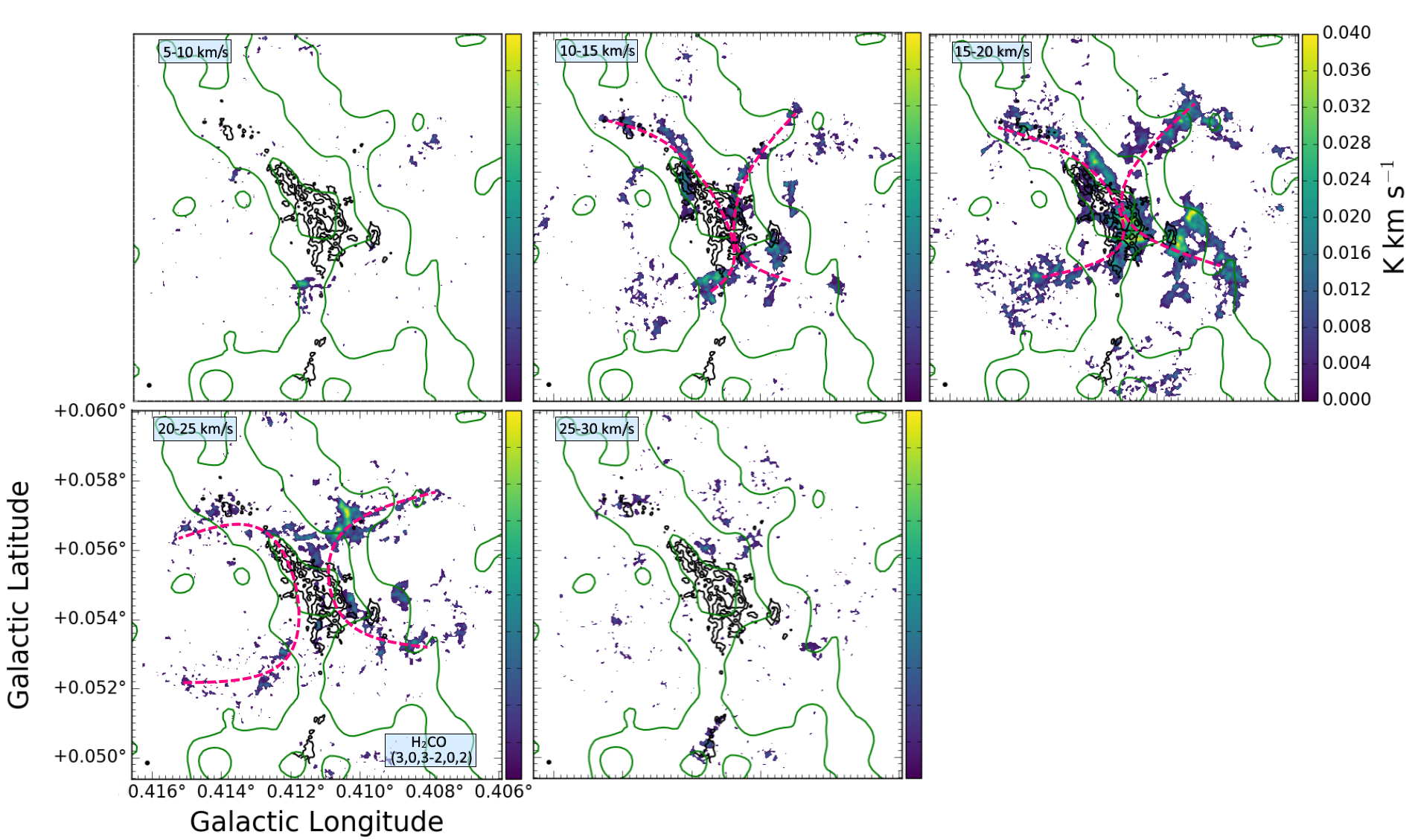}
		\caption{A channel map of H${}_2$CO (3${}_{0, 3}$-2${}_{0, 2}$) from our ALMA data in the range 5 - 30 kms${}^{-1}$. Black contours show the dust continuum emission of our ALMA data and green contours show SMA 1.3mm continuum data. The black circle in each plot represents the synthesised beam. Magenta dashed lines show the ``hollow circle" features described in $\S$~\ref{sub:cgf}. The velocity gradient in cloud `d' observed on larger scales can also be seen on this smaller scale.}
		\label{fig:h2cochan}
		\end{center}
\end{figure*}

\begin{figure*}
		\begin{center}
		\includegraphics[width=18cm]{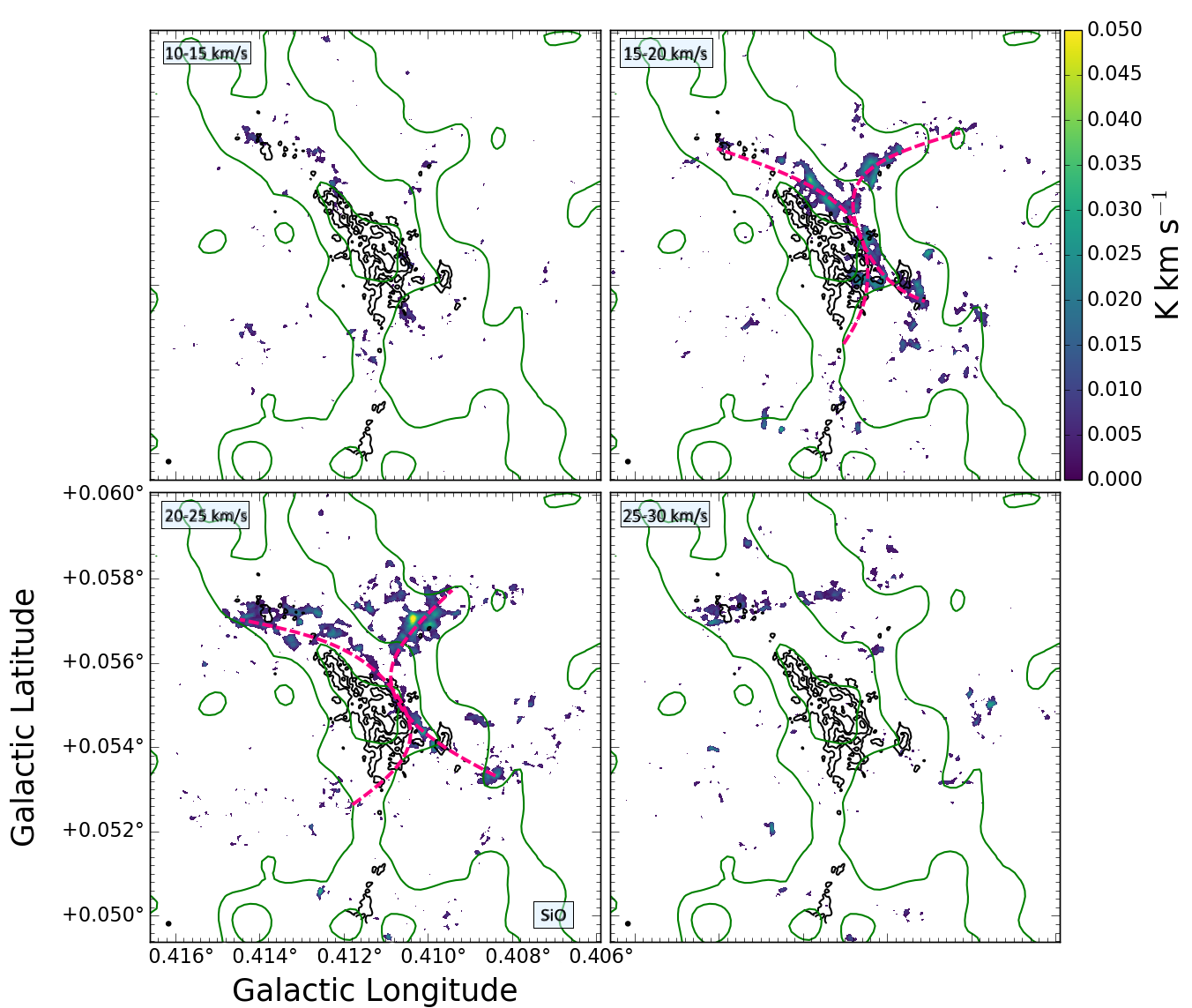}
		\caption{A channel map of SiO from our ALMA data in the range $10 - 30 $\,km\,s${}^{-1}$. Black contours show the dust continuum of our ALMA data and green contours show SMA 1.3mm continuum data. The black circle in each plot represents the synthesised beam. Magenta dashed lines show the ``hollow circle" features described in $\S$~\ref{sub:cgf}. The velocity gradient seen on larger scales in cloud `d' can be seen on this smaller scale.}
		\label{fig:siochan}
		\end{center}
\end{figure*}

To further illustrate the velocity flows we compute position-velocity (pv) diagrams of the two main flows in the HNCO Mopra data (Fig. \ref{fig:pv}). The diagrams were computed using the {\tt impv} task within {\tt CASA}. The left-hand diagram was computed with a width of 5 pixels and the right-hand diagram was computed with a width of 1 pixel. While the velocity offset is more subtle in the right-hand flow, there is still a definite offset, converging with that of the much more obvious left-hand flow.

In summary, all of the data to hand can be described as the result of a shock at the convergence point of two large-scale gas flows.



If this interpretation is correct, we estimate the time, $t$, it will take for all of the mass to end up at the convergence point through $t = d/v$, where $v$ is the velocity of the gas and $d$ is the distance the gas will travel. We use the earlier velocity dispersion of clump `d6' of $\sim$2.1 kms${}^{-1}$ from \cite{walker18}. We also use the effective radius of clump `d6' of 0.16\,pc from \cite{walker18} and double it to account for the gas moving across the entire clump. In this way, we estimate that it will take $\sim$10$^5$ years for all the mass to reach the convergent point across clump `d6'. However, clump `d6' is only a small part of cloud `d', and so we must consider the whole cloud. From \cite{walker15}, cloud `d' has a radius of 3.2 pc and line widths of $\sim$ 16\,kms${}^{-1}$, and so we estimate that all mass will converge across the entirety of cloud `d' in around 3 $\times$ 10$^{5}$ years.

We estimate the corresponding mass inflow rate, $\dot{m}$, through, $\dot{m} = m/t$, where $m$ is the gas mass. Using the upper mass limit of clump `d6' of 239 M${}_{\odot}$ from \cite{walker18}, we estimate a mass inflow rate towards the convergent point of the clump of 10$^{-3}$ M${}_{\odot}$yr${}^{-1}$. Once again considering the whole cloud, using the mass of 7.6 $\times$ 10${}^4$ M${}_{\odot}$ from \cite{walker15}, we estimate a mass inflow rate across the entire cloud of around 0.25 M${}_{\odot}$yr${}^{-1}$.

These estimates are all under the assumption that the material is all sitting in the same plane of the sky. However, there could be an offset along the line of sight, which would imply rotation, such as those in the simulations of \cite{kruijssen19}. This is because we are assuming that all of the gas is converging on one point. However, we do not know the 3D motions of the cloud. If they don't converge on this point, there could be net rotations in the gas. This may still lead to a collapse, but the process would be delayed when compared to collapse purely via convergence.

We conclude that a large reservoir of mass may be funnelled towards the convergent point in a very short period of time (although this process may be delayed if there is an offset along the line of sight). This would quickly push the gas above the critical density threshold for star formation in the CMZ \citep{walker18} and star formation will begin. This scenario would be in agreement with the results of \cite{barnes19}, such that the gas would become bound and undergo collapse following the collision of both flows. In summary, it appears that star formation is imminent in cloud `d' and that we have therefore identified a truly pre-star-forming YMC precursor gas cloud.

\subsubsection{Source of converging flows}

We now seek to understand the origin of these converging gas flows. As previously discussed, SiO emission is seen to be ``wrapping around" the right-hand edge of the dust continuum emission. The density required to excite the SiO (5-4) line across such a broad filament, coupled with the lack of SiO emission in the densest region, implies that a large-scale shock feature is present. Three plausible mechanisms which could drive large scale gas flows and generate such a shock are gravitational collapse, a cloud-cloud collision, or a shock triggered by pericenter passage with the bottom of the gravitational potential.

Cloud-cloud collisions (CCCs) -- collisions of molecular gas clouds \citep{hasegawa94} -- have been postulated as an explanation for various observed properties in multiple Galactic Centre clouds. These clouds include the `Brick' \citep{higuchi14,johnston14} and the 50 kms${}^{-1}$ cloud \citep{tsuboi15}. There are many features that are attributed to CCCs. These features include shells or cavities \citep{higuchi14,tsuboi15}, multiple velocity components connected by ``bridge features"  \citep{johnston14,tsuboi21}, and emission from shocked gas \citep{zeng20,armijos20}. Features that are unambiguously a signature of a CCC, however, are rare, particularly in an environment so complex as the CMZ. They may also occur at a much lower rate compared to other mechanisms that dominate the cloud lifetime \citep{jeffreson18}. Some shocks within the CMZ are caused by bar-driven streams colliding with clouds \citep{sormani15,sormani19,hatchfield21}. \citet{hatchfield21} find that simulated CMZ clouds have peaks in their average density at the point where they collide violently with inflowing material. However, it is unlikely that the collision observed in cloud `d' is is driven by bar inflow, as cloud `d' is not located near any proposed entry point to the CMZ \citep{henshaw22b}. It is also extremely unlikely that cloud `d' is a collision occurring downstream from an entry point, as the collision is currently ongoing. However, the shocks at cloud `d' could be the result of a collision between clouds independent of bar inflow.

Another possibility is that the shock has been caused by tidal forces in the clouds during their pericentre passage of the Galactic Centre. Hydrodynamic simulations performed by \cite{dale19} show that inclusion of tidal forces is required in modelling star formation in the centre of galaxies, and that the tidal forces experienced during pericentre passage temporarily increase the star formation rate by a factor of up to 2.7. It is possible that the shock in cloud `d' is observational evidence of the tidal deformation found in the numerical simulations by \cite{dale19}.

While these scenarios, along with pure gravitational collapse, are all possibilities for the shock we see in cloud `d', further work is required to unambiguously distinguish between them.

\section{Conclusions}

We report high-resolution (0.13$^{\prime\prime}$, 1000\,au) ALMA Band 6 (1.3\,mm, 230\,GHz) observations towards the single-dish continuum peak of the Galactic centre dust ridge cloud G0.412$+$0.052, also known as cloud `d'. We summarise the main results as follows:

\begin{itemize}
    \item This region of cloud `d' contains substructures separated on 10$^4$\,au scales. Using dendrograms to characterise the continuum structure, we identify 96 individual leaves above the 3$\sigma$ level. These  range in mass from $0.21 - 3.1$\,M$_{\odot}$, have radii of $\sim$10$^3$\,au and densities of 10$^{6-7}$\,cm$^{-3}$. Above the 5$\sigma$ level we identify nine leaves, with masses of $0.49 - 1.8$\,M$_{\odot}$ and similar radius and density ranges.
    \\
    \item The projected spatial separations of the continuum sources lie between the upper thermal Jeans length prediction and the lower turbulent Jeans length prediction. It is not clear which of these scales is favoured by the separation of the continuum sources, although there is a slight tendency towards thermal. However, the masses of the sources are consistent with thermal Jeans mass predictions but not turbulent Jeans mass predictions.
    \\
    \item The mass distribution of compact continuum sources is bottom-heavy (mean mass $\sim$0.7\,M$_{\odot}$), and does not resemble a typical stellar initial mass function or pre-stellar core mass function. Stars forming from these initial sources would produce an extremely bottom-heavy IMF. Conversely, in order to populate a normal IMF, the initial sources will have to accrete many times their current mass from the surrounding environment to form the expected large number of intermediate and high-mass stars. We expect this cloud to form many high-mass stars, but find no high-mass starless cores.
    \\
    \item None of the identified continuum sources coincide with known star formation tracers. In particular, we do not detect any molecular outflows via SiO (5-4) or $^{13}$CO (2-1) emission, nor do we detect any typical hot core tracers such as CH${}_3$CN. As star formation has been detected in similar CMZ cloud using an identical observational setup, we conclude that cloud `d' is not forming stars currently.
    \\
    \item Virial analysis suggests that the continuum sources are most likely not gravitationally bound. However, they are subject to external pressures two to three orders of magnitude greater than those found in the Galactic disc. There is either an additional confining pressure on $\sim$1000\,au scales that is undetected on the scale of clouds, or the sources are over-pressured with respect to the surrounding gas and are therefore transient. 
    \\
    \item We find evidence from single-dish molecular line observations that suggest that cloud `d' is a point of convergence of larger scale gas flows. It is estimated that all mass will have converged on this point in $\sim$10$^5$ years at a mass inflow rate of $\sim$10$^{-3}$\,M$_\odot$\,yr$^{-1}$. If the cloud continues to collapse and the sources continue to grow via accretion of this material, then the cloud has the potential to begin forming stars in this time frame, although we cannot confirm the mechanism by which they will form.
\end{itemize}

We conclude that cloud `d' is the earliest known pre-star forming massive cluster and therefore is an ideal laboratory in which to study the initial conditions of star and cluster formation in extreme environments. These initial conditions shape the IMF and set global star forming relations in extreme (but cosmologically typical) conditions and therefore further study of cloud `d' and its counterparts is important to further our understanding.

\section*{Acknowledgements}
BAW thanks Matt, Luke and Ash for their continued support and encouragement, and acknowledges an STFC doctoral studentship. DLW and CB acknowledge support from the National Science Foundation under Award No. 1816715 and CB also acknowledges support under Award No. 2108938. ATB would like to acknowledge funding from the European Research Council (ERC) under the European Union's Horizon 2020 research and innovation programme (grant agreement No.726384/Empire). G.G. acknowledges support from ANID project FB210003. LCH  was supported by the National Science Foundation of China (11721303, 11991052) and the National Key R\&D Program of China (2016YFA0400702). JMDK and MAP gratefully acknowledge funding from the European Research Council (ERC) under the European Union's Horizon 2020 research and innovation programme via the ERC Starting Grant MUSTANG (grant agreement number 714907). JMDK gratefully acknowledges funding from the Deutsche Forschungsgemeinschaft (DFG) in the form of an Emmy Noether Research Group (grant number KR4801/1-1). XL was supported by JSPS KAKENHI grant No.\ 20K14528.

{\bf Data availability}: The data products used to conduct the research presented in this paper are made publicly available at the following Zenodo repository: \url{https://doi.org/10.5281/zenodo.6546462}. Scripts used to conduct this research are also made available on GitHub at: \url{https://github.com/b-a-williams/Cloud-d-ALMA}.





\bibliographystyle{mnras}
\bibliography{bibliography}

\begin{thebibliography}{}
\makeatletter
\relax
\def\mn@urlcharsother{\let\do\@makeother \do\$\do\&\do\#\do\^\do\_\do\%\do\~}
\def\mn@doi{\begingroup\mn@urlcharsother \@ifnextchar [ {\mn@doi@}
  {\mn@doi@[]}}
\def\mn@doi@[#1]#2{\def\@tempa{#1}\ifx\@tempa\@empty \href
  {http://dx.doi.org/#2} {doi:#2}\else \href {http://dx.doi.org/#2} {#1}\fi
  \endgroup}
\def\mn@eprint#1#2{\mn@eprint@#1:#2::\@nil}
\def\mn@eprint@arXiv#1{\href {http://arxiv.org/abs/#1} {{\tt arXiv:#1}}}
\def\mn@eprint@dblp#1{\href {http://dblp.uni-trier.de/rec/bibtex/#1.xml}
  {dblp:#1}}
\def\mn@eprint@#1:#2:#3:#4\@nil{\def\@tempa {#1}\def\@tempb {#2}\def\@tempc
  {#3}\ifx \@tempc \@empty \let \@tempc \@tempb \let \@tempb \@tempa \fi \ifx
  \@tempb \@empty \def\@tempb {arXiv}\fi \@ifundefined
  {mn@eprint@\@tempb}{\@tempb:\@tempc}{\expandafter \expandafter \csname
  mn@eprint@\@tempb\endcsname \expandafter{\@tempc}}}

\bibitem[\protect\citeauthoryear{{Abuter} et~al.,}{{Abuter}
  et~al.}{2019}]{abuter19}
{Abuter} R.,  et~al., 2019, \aap, 625, L10

\bibitem[\protect\citeauthoryear{{Armijos-Abenda{\~n}o}, {Banda-Barrag{\'a}n},
  {Mart{\'\i}n-Pintado}, {D{\'e}nes}, {Federrath}  \&
  {Requena-Torres}}{{Armijos-Abenda{\~n}o} et~al.}{2020}]{armijos20}
{Armijos-Abenda{\~n}o} J.,  {Banda-Barrag{\'a}n} W.~E.,  {Mart{\'\i}n-Pintado}
  J.,  {D{\'e}nes} H.,  {Federrath} C.,   {Requena-Torres} M.~A.,  2020,
  \mn@doi [\mnras] {10.1093/mnras/staa3119}, \href
  {https://ui.adsabs.harvard.edu/abs/2020MNRAS.499.4918A} {499, 4918}

\bibitem[\protect\citeauthoryear{{Bally}}{{Bally}}{2016}]{bally16}
{Bally} J.,  2016, \mn@doi [\araa] {10.1146/annurev-astro-081915-023341}, \href
  {https://ui.adsabs.harvard.edu/abs/2016ARA&A..54..491B} {54, 491}

\bibitem[\protect\citeauthoryear{{Barnes} et~al.,}{{Barnes}
  et~al.}{2019}]{barnes19}
{Barnes} A.~T.,  et~al., 2019, \mnras, 486, 283

\bibitem[\protect\citeauthoryear{{Battersby} et~al.,}{{Battersby}
  et~al.}{2011}]{battersby11}
{Battersby} C.,  et~al., 2011, \aap, 535, A128

\bibitem[\protect\citeauthoryear{{Battersby} et~al.,}{{Battersby}
  et~al.}{2020}]{battersby20}
{Battersby} C.,  et~al., 2020, \apjs, 249, 35

\bibitem[\protect\citeauthoryear{{Bertoldi} \& {McKee}}{{Bertoldi} \&
  {McKee}}{1992}]{bertoldi92}
{Bertoldi} F.,  {McKee} C.~F.,  1992, \mn@doi [\apj] {10.1086/171638}, \href
  {https://ui.adsabs.harvard.edu/abs/1992ApJ...395..140B} {395, 140}

\bibitem[\protect\citeauthoryear{{Beuther}, {Walsh}, {Johnston}, {Henning},
  {Kuiper}, {Longmore}  \& {Walmsley}}{{Beuther} et~al.}{2017}]{beuther17}
{Beuther} H.,  {Walsh} A.~J.,  {Johnston} K.~G.,  {Henning} T.,  {Kuiper} R.,
  {Longmore} S.~N.,   {Walmsley} C.~M.,  2017, \mn@doi [\aap]
  {10.1051/0004-6361/201630126}, \href
  {https://ui.adsabs.harvard.edu/abs/2017A&A...603A..10B} {603, A10}

\bibitem[\protect\citeauthoryear{{Bressert}, {Ginsburg}, {Bally}, {Battersby},
  {Longmore}  \& {Testi}}{{Bressert} et~al.}{2012}]{bressert12}
{Bressert} E.,  {Ginsburg} A.,  {Bally} J.,  {Battersby} C.,  {Longmore} S.,
  {Testi} L.,  2012, \mn@doi [\apjl] {10.1088/2041-8205/758/2/L28}, \href
  {https://ui.adsabs.harvard.edu/abs/2012ApJ...758L..28B} {758, L28}

\bibitem[\protect\citeauthoryear{Churchwell et~al.,}{Churchwell
  et~al.}{2009}]{churchwell09}
Churchwell E.,  et~al., 2009, Publications of the Astronomical Society of the
  Pacific, 121, 213

\bibitem[\protect\citeauthoryear{{Clark}, {Glover}, {Ragan}, {Shetty}  \&
  {Klessen}}{{Clark} et~al.}{2013}]{clark2013}
{Clark} P.~C.,  {Glover} S. C.~O.,  {Ragan} S.~E.,  {Shetty} R.,   {Klessen}
  R.~S.,  2013, \mn@doi [\apjl] {10.1088/2041-8205/768/2/L34}, \href
  {https://ui.adsabs.harvard.edu/abs/2013ApJ...768L..34C} {768, L34}

\bibitem[\protect\citeauthoryear{{Contreras}, {Rathborne}, {Guzman}, {Jackson},
  {Whitaker}, {Sanhueza}  \& {Foster}}{{Contreras} et~al.}{2017}]{contreras17}
{Contreras} Y.,  {Rathborne} J.~M.,  {Guzman} A.,  {Jackson} J.,  {Whitaker}
  S.,  {Sanhueza} P.,   {Foster} J.,  2017, \mn@doi [\mnras]
  {10.1093/mnras/stw3110}, \href
  {https://ui.adsabs.harvard.edu/abs/2017MNRAS.466..340C} {466, 340}

\bibitem[\protect\citeauthoryear{{Cotton} \& {Yusef-Zadeh}}{{Cotton} \&
  {Yusef-Zadeh}}{2016}]{cotton16}
{Cotton} W.~D.,  {Yusef-Zadeh} F.,  2016, \apjs, 227, 10

\bibitem[\protect\citeauthoryear{{Dale}, {Kruijssen}  \& {Longmore}}{{Dale}
  et~al.}{2019a}]{dkl19}
{Dale} J.~E.,  {Kruijssen} J.~M.~D.,   {Longmore} S.~N.,  2019a, \mn@doi
  [\mnras] {10.1093/mnras/stz888}, \href
  {https://ui.adsabs.harvard.edu/abs/2019MNRAS.486.3307D} {486, 3307}

\bibitem[\protect\citeauthoryear{{Dale}, {Kruijssen}  \& {Longmore}}{{Dale}
  et~al.}{2019b}]{dale19}
{Dale} J.~E.,  {Kruijssen} J.~M.~D.,   {Longmore} S.~N.,  2019b, \mn@doi
  [\mnras] {10.1093/mnras/stz888}, \href
  {https://ui.adsabs.harvard.edu/abs/2019MNRAS.486.3307D} {486, 3307}

\bibitem[\protect\citeauthoryear{{Elmegreen}}{{Elmegreen}}{1989}]{elmegreen89}
{Elmegreen} B.~G.,  1989, \apj, 338, 178

\bibitem[\protect\citeauthoryear{{Elmegreen} \& {Efremov}}{{Elmegreen} \&
  {Efremov}}{1997}]{elmegreen97}
{Elmegreen} B.~G.,  {Efremov} Y.~N.,  1997, \apj, 480, 235

\bibitem[\protect\citeauthoryear{{Field}, {Blackman}  \& {Keto}}{{Field}
  et~al.}{2011}]{field11}
{Field} G.~B.,  {Blackman} E.~G.,   {Keto} E.~R.,  2011, \mnras, 416, 710

\bibitem[\protect\citeauthoryear{{Foster} et~al.,}{{Foster}
  et~al.}{2011}]{foster11}
{Foster} J.~B.,  et~al., 2011, \apjs, 197, 25

\bibitem[\protect\citeauthoryear{{Foster} et~al.,}{{Foster}
  et~al.}{2013}]{foster13}
{Foster} J.~B.,  et~al., 2013, \pasa, 30, e038

\bibitem[\protect\citeauthoryear{{Giannetti} et~al.,}{{Giannetti}
  et~al.}{2017}]{giannetti17}
{Giannetti} A.,  et~al., 2017, \mn@doi [\aap] {10.1051/0004-6361/201731728},
  \href {https://ui.adsabs.harvard.edu/abs/2017A&A...606L..12G} {606, L12}

\bibitem[\protect\citeauthoryear{{Ginsburg} \& {Kruijssen}}{{Ginsburg} \&
  {Kruijssen}}{2018}]{ginsburg18b}
{Ginsburg} A.,  {Kruijssen} J.~M.~D.,  2018, \mn@doi [\apjl]
  {10.3847/2041-8213/aada89}, \href
  {https://ui.adsabs.harvard.edu/abs/2018ApJ...864L..17G} {864, L17}

\bibitem[\protect\citeauthoryear{{Ginsburg}, {Bressert}, {Bally}  \&
  {Battersby}}{{Ginsburg} et~al.}{2012}]{ginsburg12}
{Ginsburg} A.,  {Bressert} E.,  {Bally} J.,   {Battersby} C.,  2012, \mn@doi
  [\apjl] {10.1088/2041-8205/758/2/L29}, \href
  {https://ui.adsabs.harvard.edu/abs/2012ApJ...758L..29G} {758, L29}

\bibitem[\protect\citeauthoryear{{Ginsburg} et~al.,}{{Ginsburg}
  et~al.}{2013}]{ginsburg13}
{Ginsburg} A.,  et~al., 2013, \mn@doi [\apjs] {10.1088/0067-0049/208/2/14},
  \href {https://ui.adsabs.harvard.edu/abs/2013ApJS..208...14G} {208, 14}

\bibitem[\protect\citeauthoryear{{Ginsburg} et~al.,}{{Ginsburg}
  et~al.}{2016}]{ginsburg16}
{Ginsburg} A.,  et~al., 2016, \mn@doi [\aap] {10.1051/0004-6361/201526100},
  \href {https://ui.adsabs.harvard.edu/abs/2016A&A...586A..50G} {586, A50}

\bibitem[\protect\citeauthoryear{{Ginsburg} et~al.,}{{Ginsburg}
  et~al.}{2018}]{ginsburg18}
{Ginsburg} A.,  et~al., 2018, \mn@doi [\apj] {10.3847/1538-4357/aaa6d4}, \href
  {https://ui.adsabs.harvard.edu/abs/2018ApJ...853..171G} {853, 171}

\bibitem[\protect\citeauthoryear{Guesten \& Henkel}{Guesten \&
  Henkel}{1983}]{guesten83}
Guesten R.,  Henkel C.,  1983, \aap, 125, 136

\bibitem[\protect\citeauthoryear{{Hasegawa}, {Sato}, {Whiteoak}  \&
  {Miyawaki}}{{Hasegawa} et~al.}{1994}]{hasegawa94}
{Hasegawa} T.,  {Sato} F.,  {Whiteoak} J.~B.,   {Miyawaki} R.,  1994, \mn@doi
  [\apjl] {10.1086/187417}, \href
  {https://ui.adsabs.harvard.edu/abs/1994ApJ...429L..77H} {429, L77}

\bibitem[\protect\citeauthoryear{{Hatchfield} et~al.,}{{Hatchfield}
  et~al.}{2020}]{hatchfield20}
{Hatchfield} H.~P.,  et~al., 2020, \mn@doi [\apjs] {10.3847/1538-4365/abb610},
  \href {https://ui.adsabs.harvard.edu/abs/2020ApJS..251...14H} {251, 14}

\bibitem[\protect\citeauthoryear{{Hatchfield}, {Sormani}, {Tress}, {Battersby},
  {Smith}, {Glover}  \& {Klessen}}{{Hatchfield} et~al.}{2021}]{hatchfield21}
{Hatchfield} H.~P.,  {Sormani} M.~C.,  {Tress} R.~G.,  {Battersby} C.,  {Smith}
  R.~J.,  {Glover} S. C.~O.,   {Klessen} R.~S.,  2021, \mn@doi [\apj]
  {10.3847/1538-4357/ac1e89}, \href
  {https://ui.adsabs.harvard.edu/abs/2021ApJ...922...79H} {922, 79}

\bibitem[\protect\citeauthoryear{{Henshaw} et~al.,}{{Henshaw}
  et~al.}{2016a}]{henshaw16}
{Henshaw} J.~D.,  et~al., 2016a, \mn@doi [\mnras] {10.1093/mnras/stw121}, \href
  {https://ui.adsabs.harvard.edu/abs/2016MNRAS.457.2675H} {457, 2675}

\bibitem[\protect\citeauthoryear{{Henshaw} et~al.,}{{Henshaw}
  et~al.}{2016b}]{henshaw16b}
{Henshaw} J.~D.,  et~al., 2016b, \mn@doi [\mnras] {10.1093/mnras/stw1794},
  \href {https://ui.adsabs.harvard.edu/abs/2016MNRAS.463..146H} {463, 146}

\bibitem[\protect\citeauthoryear{{Henshaw}, {Longmore}  \&
  {Kruijssen}}{{Henshaw} et~al.}{2016c}]{henshaw16c}
{Henshaw} J.~D.,  {Longmore} S.~N.,   {Kruijssen} J.~M.~D.,  2016c, \mn@doi
  [\mnras] {10.1093/mnrasl/slw168}, \href
  {https://ui.adsabs.harvard.edu/abs/2016MNRAS.463L.122H} {463, L122}

\bibitem[\protect\citeauthoryear{{Henshaw} et~al.,}{{Henshaw}
  et~al.}{2019}]{henshaw19}
{Henshaw} J.~D.,  et~al., 2019, \mn@doi [\mnras] {10.1093/mnras/stz471}, \href
  {https://ui.adsabs.harvard.edu/abs/2019MNRAS.485.2457H} {485, 2457}

\bibitem[\protect\citeauthoryear{{Henshaw}, {Barnes}, {Battersby}, {Ginsburg},
  {Sormani}  \& {Walker}}{{Henshaw} et~al.}{2022a}]{henshaw22b}
{Henshaw} J.~D.,  {Barnes} A.~T.,  {Battersby} C.,  {Ginsburg} A.,  {Sormani}
  M.~C.,   {Walker} D.~L.,  2022a, arXiv e-prints, \href
  {https://ui.adsabs.harvard.edu/abs/2022arXiv220311223H} {p. arXiv:2203.11223}

\bibitem[\protect\citeauthoryear{{Henshaw} et~al.,}{{Henshaw}
  et~al.}{2022b}]{henshaw22a}
{Henshaw} J.~D.,  et~al., 2022b, \mn@doi [\mnras] {10.1093/mnras/stab3039},
  \href {https://ui.adsabs.harvard.edu/abs/2022MNRAS.509.4758H} {509, 4758}

\bibitem[\protect\citeauthoryear{{Heyer}, {Krawczyk}, {Duval}  \&
  {Jackson}}{{Heyer} et~al.}{2009}]{heyer09}
{Heyer} M.,  {Krawczyk} C.,  {Duval} J.,   {Jackson} J.~M.,  2009, \apj, 699,
  1092

\bibitem[\protect\citeauthoryear{{Higuchi}, {Chibueze}, {Habe}, {Takahira}  \&
  {Takano}}{{Higuchi} et~al.}{2014}]{higuchi14}
{Higuchi} A.~E.,  {Chibueze} J.~O.,  {Habe} A.,  {Takahira} K.,   {Takano} S.,
  2014, \mn@doi [\aj] {10.1088/0004-6256/147/6/141}, \href
  {https://ui.adsabs.harvard.edu/abs/2014AJ....147..141H} {147, 141}

\bibitem[\protect\citeauthoryear{{Ilee}, {Cyganowski}, {Brogan}, {Hunter},
  {Forgan}, {Haworth}, {Clarke}  \& {Harries}}{{Ilee} et~al.}{2018}]{ilee18}
{Ilee} J.~D.,  {Cyganowski} C.~J.,  {Brogan} C.~L.,  {Hunter} T.~R.,  {Forgan}
  D.~H.,  {Haworth} T.~J.,  {Clarke} C.~J.,   {Harries} T.~J.,  2018, \mn@doi
  [\apjl] {10.3847/2041-8213/aaeffc}, \href
  {https://ui.adsabs.harvard.edu/abs/2018ApJ...869L..24I} {869, L24}

\bibitem[\protect\citeauthoryear{{Immer}, Menten, Schuller  \& Lis}{{Immer}
  et~al.}{2012}]{immer12}
{Immer} K.,  Menten K.~M.,  Schuller F.,   Lis D.~C.,  2012, \aap, 548, A120

\bibitem[\protect\citeauthoryear{{Immer}, {Kauffmann}, {Pillai}, {Ginsburg}  \&
  {Menten}}{{Immer} et~al.}{2016}]{immer16}
{Immer} K.,  {Kauffmann} J.,  {Pillai} T.,  {Ginsburg} A.,   {Menten} K.~M.,
  2016, \aap, 595, A94

\bibitem[\protect\citeauthoryear{{Jackson} et~al.,}{{Jackson}
  et~al.}{2006}]{jackson06}
{Jackson} J.~M.,  et~al., 2006, \mn@doi [\apjs] {10.1086/500091}, \href
  {https://ui.adsabs.harvard.edu/abs/2006ApJS..163..145J} {163, 145}

\bibitem[\protect\citeauthoryear{{Jackson} et~al.,}{{Jackson}
  et~al.}{2013}]{jackson13}
{Jackson} J.~M.,  et~al., 2013, \pasa, 30, e057

\bibitem[\protect\citeauthoryear{{Jackson} et~al.,}{{Jackson}
  et~al.}{2018}]{jackson18}
{Jackson} J.~M.,  et~al., 2018, \mn@doi [\apj] {10.3847/1538-4357/aae7c7},
  \href {https://ui.adsabs.harvard.edu/abs/2018ApJ...869..102J} {869, 102}

\bibitem[\protect\citeauthoryear{{Jeffreson}, {Kruijssen}, {Krumholz}  \&
  {Longmore}}{{Jeffreson} et~al.}{2018}]{jeffreson18}
{Jeffreson} S.~M.~R.,  {Kruijssen} J.~M.~D.,  {Krumholz} M.~R.,   {Longmore}
  S.~N.,  2018, \mn@doi [\mnras] {10.1093/mnras/sty1154}, \href
  {https://ui.adsabs.harvard.edu/abs/2018MNRAS.478.3380J} {478, 3380}

\bibitem[\protect\citeauthoryear{{Johnston}, {Beuther}, {Linz}, {Schmiedeke},
  {Ragan}  \& {Henning}}{{Johnston} et~al.}{2014}]{johnston14}
{Johnston} K.~G.,  {Beuther} H.,  {Linz} H.,  {Schmiedeke} A.,  {Ragan} S.~E.,
   {Henning} T.,  2014, \mn@doi [\aap] {10.1051/0004-6361/201423943}, \href
  {https://ui.adsabs.harvard.edu/abs/2014A&A...568A..56J} {568, A56}

\bibitem[\protect\citeauthoryear{{Kauffmann} \& {Pillai}}{{Kauffmann} \&
  {Pillai}}{2010}]{kauffmann10}
{Kauffmann} J.,  {Pillai} T.,  2010, \apjl, 723, L7

\bibitem[\protect\citeauthoryear{{Kauffmann}, Bertoldi, Bourke, Evans~II  \&
  Lee}{{Kauffmann} et~al.}{2008}]{kauffmann08}
{Kauffmann} J.,  Bertoldi F.,  Bourke T.~L.,  Evans~II N.~J.,   Lee C.~W.,
  2008, {\aap}, 487, 993

\bibitem[\protect\citeauthoryear{{Krieger} et~al.,}{{Krieger}
  et~al.}{2017}]{krieger17}
{Krieger} N.,  et~al., 2017, \apj, 850, 77

\bibitem[\protect\citeauthoryear{{Kruijssen}}{{Kruijssen}}{2015}]{kruijssen15b}
{Kruijssen} J.~M.~D.,  2015, \mnras, 454, 1658

\bibitem[\protect\citeauthoryear{Kruijssen, Longmore, Elmegreen, Murray, Bally,
  Testi  \& Kennicutt}{Kruijssen et~al.}{2014}]{kruijssen14}
Kruijssen J. M.~D.,  Longmore S.~N.,  Elmegreen B.~G.,  Murray N.,  Bally J.,
  Testi L.,   Kennicutt R.~C.,  2014, \mnras, 440, 3370

\bibitem[\protect\citeauthoryear{{Kruijssen}, {Dale}  \&
  {Longmore}}{{Kruijssen} et~al.}{2015}]{kruijssen15}
{Kruijssen} J.~M.~D.,  {Dale} J.~E.,   {Longmore} S.~N.,  2015, \mnras, 447,
  1059

\bibitem[\protect\citeauthoryear{{Kruijssen} et~al.,}{{Kruijssen}
  et~al.}{2019a}]{kdlplus19}
{Kruijssen} J.~M.~D.,  et~al., 2019a, \mn@doi [\mnras] {10.1093/mnras/stz381},
  \href {https://ui.adsabs.harvard.edu/abs/2019MNRAS.484.5734K} {484, 5734}

\bibitem[\protect\citeauthoryear{{Kruijssen} et~al.,}{{Kruijssen}
  et~al.}{2019b}]{kruijssen19}
{Kruijssen} J.~M.~D.,  et~al., 2019b, \mn@doi [\mnras] {10.1093/mnras/stz381},
  \href {https://ui.adsabs.harvard.edu/abs/2019MNRAS.484.5734K} {484, 5734}

\bibitem[\protect\citeauthoryear{{Kruijssen} et~al.,}{{Kruijssen}
  et~al.}{2019c}]{kruijssen19b}
{Kruijssen} J.~M.~D.,  et~al., 2019c, \mn@doi [\nat]
  {10.1038/s41586-019-1194-3}, \href
  {https://ui.adsabs.harvard.edu/abs/2019Natur.569..519K} {569, 519}

\bibitem[\protect\citeauthoryear{{Krumholz} \& {McKee}}{{Krumholz} \&
  {McKee}}{2020}]{krumholz20}
{Krumholz} M.~R.,  {McKee} C.~F.,  2020, \mnras, 494, 624

\bibitem[\protect\citeauthoryear{{Krumholz}, Klein  \& McKee"}{{Krumholz}
  et~al.}{2007}]{krumholz07}
{Krumholz} M.~R.,  Klein R.~I.,   McKee" C.~F.,  2007, \apj, 665, 478

\bibitem[\protect\citeauthoryear{{Lada}, {Lombardi}  \& {Alves}}{{Lada}
  et~al.}{2010}]{lada10}
{Lada} C.~J.,  {Lombardi} M.,   {Alves} J.~F.,  2010, \apj, 724, 687

\bibitem[\protect\citeauthoryear{{Lis}, {Menten}, {Serabyn}  \& {Zylka}}{{Lis}
  et~al.}{1994}]{lis94}
{Lis} D.~C.,  {Menten} K.~M.,  {Serabyn} E.,   {Zylka} R.,  1994, \mn@doi
  [\apjl] {10.1086/187230}, \href
  {https://ui.adsabs.harvard.edu/abs/1994ApJ...423L..39L} {423, L39}

\bibitem[\protect\citeauthoryear{{Longmore} et~al.,}{{Longmore}
  et~al.}{2012}]{longmore12}
{Longmore} S.~N.,  et~al., 2012, \mn@doi [\apj] {10.1088/0004-637X/746/2/117},
  \href {https://ui.adsabs.harvard.edu/abs/2012ApJ...746..117L} {746, 117}

\bibitem[\protect\citeauthoryear{{Longmore} et~al.,}{{Longmore}
  et~al.}{2013a}]{longmore13}
{Longmore} S.~N.,  et~al., 2013a, \mnras, 429, 987

\bibitem[\protect\citeauthoryear{{Longmore} et~al.,}{{Longmore}
  et~al.}{2013b}]{longmore13b}
{Longmore} S.~N.,  et~al., 2013b, \mnras, 433, L15

\bibitem[\protect\citeauthoryear{{Longmore} et~al.,}{{Longmore}
  et~al.}{2014}]{longmore14}
{Longmore} S.~N.,  et~al., 2014, in {Beuther} H.,  {Klessen} R.~S.,
  {Dullemond} C.~P.,   {Henning} T.,  eds, Protostars and Planets VI. p.~291

\bibitem[\protect\citeauthoryear{{Longmore} et~al.,}{{Longmore}
  et~al.}{2017}]{longmore17}
{Longmore} S.~N.,  et~al., 2017, \mnras, 470, 1462

\bibitem[\protect\citeauthoryear{{Louvet} et~al.,}{{Louvet}
  et~al.}{2014}]{louvet14}
{Louvet} F.,  et~al., 2014, \aap, 570, A15

\bibitem[\protect\citeauthoryear{{Lu} et~al.,}{{Lu} et~al.}{2019a}]{lu19b}
{Lu} X.,  et~al., 2019a, \apjs, 244, 35

\bibitem[\protect\citeauthoryear{{Lu} et~al.,}{{Lu} et~al.}{2019b}]{lu19}
{Lu} X.,  et~al., 2019b, \apj, 872, 171

\bibitem[\protect\citeauthoryear{{Lu}, {Cheng}, {Ginsburg}, {Longmore},
  {Kruijssen}, {Battersby}, {Zhang}  \& {Walker}}{{Lu} et~al.}{2020}]{lu20}
{Lu} X.,  {Cheng} Y.,  {Ginsburg} A.,  {Longmore} S.~N.,  {Kruijssen} J.~M.~D.,
   {Battersby} C.,  {Zhang} Q.,   {Walker} D.~L.,  2020, \apjl, 894, L14

\bibitem[\protect\citeauthoryear{{Lu} et~al.,}{{Lu} et~al.}{2021}]{lu21}
{Lu} X.,  et~al., 2021, \apj, 909, 177

\bibitem[\protect\citeauthoryear{{Marsh} et~al.,}{{Marsh}
  et~al.}{2017}]{marsh17}
{Marsh} K.~A.,  et~al., 2017, \mnras, 471, 2730

\bibitem[\protect\citeauthoryear{{Maud} et~al.,}{{Maud} et~al.}{2018}]{maud18}
{Maud} L.~T.,  et~al., 2018, \mn@doi [\aap] {10.1051/0004-6361/201833908},
  \href {https://ui.adsabs.harvard.edu/abs/2018A&A...620A..31M} {620, A31}

\bibitem[\protect\citeauthoryear{{McMullin}, Waters, Schiebel, Young  \&
  Golap}{{McMullin} et~al.}{2007}]{mcmullin07}
{McMullin} J.~P.,  Waters B.,  Schiebel D.,  Young W.,   Golap K.,  2007, in
  Shaw R.~A.,  Hill F.,   Bell D.~J.,  eds, , ASP Conference Series.
Astron. Soc. Pac., San Francisco, p.~127

\bibitem[\protect\citeauthoryear{Mills \& Morris}{Mills \&
  Morris}{2013}]{mills13}
Mills E. A.~C.,  Morris M.~R.,  2013, \apj, 772, 105

\bibitem[\protect\citeauthoryear{{Mills}, {Butterfield}, {Ludovici}, {Lang},
  {Ott}, {Morris}  \& {Schmitz}}{{Mills} et~al.}{2015}]{mills15}
{Mills} E.~A.~C.,  {Butterfield} N.,  {Ludovici} D.~A.,  {Lang} C.~C.,  {Ott}
  J.,  {Morris} M.~R.,   {Schmitz} S.,  2015, \mn@doi [\apj]
  {10.1088/0004-637X/805/1/72}, \href
  {https://ui.adsabs.harvard.edu/abs/2015ApJ...805...72M} {805, 72}

\bibitem[\protect\citeauthoryear{{Mills}, {Ginsburg}, {Immer}, {Barnes},
  {Wiesenfeld}, {Faure}, {Morris}  \& {Requena-Torres}}{{Mills}
  et~al.}{2018}]{mills18}
{Mills} E.~A.~C.,  {Ginsburg} A.,  {Immer} K.,  {Barnes} J.~M.,  {Wiesenfeld}
  L.,  {Faure} A.,  {Morris} M.~R.,   {Requena-Torres} M.~A.,  2018, \mn@doi
  [\apj] {10.3847/1538-4357/aae581}, \href
  {https://ui.adsabs.harvard.edu/abs/2018ApJ...868....7M} {868, 7}

\bibitem[\protect\citeauthoryear{Molinari, Swinyard, Bally, Barlow  \&
  Bernard}{Molinari et~al.}{2010}]{molinari10}
Molinari S.,  Swinyard B.,  Bally J.,  Barlow M.,   Bernard J.,  2010,
  Publications of the Astronomical Society of the Pacific, 122, 314

\bibitem[\protect\citeauthoryear{{Ossenkopf} \& {Henning}}{{Ossenkopf} \&
  {Henning}}{1994}]{ossenkopf94}
{Ossenkopf} V.,  {Henning} T.,  1994, \aap, \href
  {https://ui.adsabs.harvard.edu/abs/1994A&A...291..943O} {291, 943}

\bibitem[\protect\citeauthoryear{{Padoan} \& {Nordlund}}{{Padoan} \&
  {Nordlund}}{2011}]{padoan11}
{Padoan} P.,  {Nordlund} {\r{A}}.,  2011, in {Alves} J.,  {Elmegreen} B.~G.,
  {Girart} J.~M.,   {Trimble} V.,  eds, ~1 Vol. 270, Computational Star
  Formation. pp 347--354, \mn@doi{10.1017/S1743921311000615}

\bibitem[\protect\citeauthoryear{{Peretto} et~al.,}{{Peretto}
  et~al.}{2013}]{peretto13}
{Peretto} N.,  et~al., 2013, \aap, 555, A112

\bibitem[\protect\citeauthoryear{{Petkova} et~al.,}{{Petkova}
  et~al.}{2021}]{petkova21}
{Petkova} M.~A.,  et~al., 2021, arXiv e-prints, \href
  {https://ui.adsabs.harvard.edu/abs/2021arXiv210409558P} {p. arXiv:2104.09558}

\bibitem[\protect\citeauthoryear{{Pfeffer}, {Kruijssen}, {Crain}  \&
  {Bastian}}{{Pfeffer} et~al.}{2018}]{pfeffer18}
{Pfeffer} J.,  {Kruijssen} J.~M.~D.,  {Crain} R.~A.,   {Bastian} N.,  2018,
  \mn@doi [\mnras] {10.1093/mnras/stx3124}, \href
  {https://ui.adsabs.harvard.edu/abs/2018MNRAS.475.4309P} {475, 4309}

\bibitem[\protect\citeauthoryear{{Pineda} et~al.,}{{Pineda}
  et~al.}{2015}]{pineda15}
{Pineda} J.~E.,  et~al., 2015, \mn@doi [\nat] {10.1038/nature14166}, \href
  {https://ui.adsabs.harvard.edu/abs/2015Natur.518..213P} {518, 213}

\bibitem[\protect\citeauthoryear{{Portegies Zwart}, {McMillan}  \&
  {Gieles}}{{Portegies Zwart} et~al.}{2010}]{portegies10}
{Portegies Zwart} S.~F.,  {McMillan} S. L.~W.,   {Gieles} M.,  2010, \araa, 48,
  431

\bibitem[\protect\citeauthoryear{{Rathborne} et~al.,}{{Rathborne}
  et~al.}{2014}]{rathborne14}
{Rathborne} J.~M.,  et~al., 2014, \apj, 795, L25

\bibitem[\protect\citeauthoryear{{Rathborne} et~al.,}{{Rathborne}
  et~al.}{2015}]{rathborne15}
{Rathborne} J.~M.,  et~al., 2015, \apj, 802, 125

\bibitem[\protect\citeauthoryear{{Rickert}, {Yusef-Zadeh}  \& {Ott}}{{Rickert}
  et~al.}{2019}]{rickert19}
{Rickert} M.,  {Yusef-Zadeh} F.,   {Ott} J.,  2019, \mnras, 482, 5349

\bibitem[\protect\citeauthoryear{{Rodr{\'\i}guez} \& {Zapata}}{{Rodr{\'\i}guez}
  \& {Zapata}}{2013}]{rodriguez13}
{Rodr{\'\i}guez} L.~F.,  {Zapata} L.~A.,  2013, \mn@doi [\apjl]
  {10.1088/2041-8205/767/1/L13}, \href
  {https://ui.adsabs.harvard.edu/abs/2013ApJ...767L..13R} {767, L13}

\bibitem[\protect\citeauthoryear{{Rosolowsky} \& {Leroy}}{{Rosolowsky} \&
  {Leroy}}{2006}]{rosolowsky06}
{Rosolowsky} E.,  {Leroy} A.,  2006, \mn@doi [\pasp] {10.1086/502982}, \href
  {https://ui.adsabs.harvard.edu/abs/2006PASP..118..590R} {118, 590}

\bibitem[\protect\citeauthoryear{{Rosolowsky}, {Pineda}, {Kauffmann}  \&
  {Goodman}}{{Rosolowsky} et~al.}{2008}]{rosolowsky08}
{Rosolowsky} E.~W.,  {Pineda} J.~E.,  {Kauffmann} J.,   {Goodman} A.~A.,  2008,
  \apj, 679, 1338

\bibitem[\protect\citeauthoryear{{Salpeter}}{{Salpeter}}{1955}]{salpeter55}
{Salpeter} E.~E.,  1955, \mn@doi [\apj] {10.1086/145971}, \href
  {https://ui.adsabs.harvard.edu/abs/1955ApJ...121..161S} {121, 161}

\bibitem[\protect\citeauthoryear{{Schw{\"o}rer} et~al.,}{{Schw{\"o}rer}
  et~al.}{2019}]{schworer19}
{Schw{\"o}rer} A.,  et~al., 2019, \mn@doi [\aap] {10.1051/0004-6361/201935200},
  \href {https://ui.adsabs.harvard.edu/abs/2019A&A...628A...6S} {628, A6}

\bibitem[\protect\citeauthoryear{{Shetty}, {Beaumont}, {Burton}, {Kelly}  \&
  {Klessen}}{{Shetty} et~al.}{2012}]{shetty12}
{Shetty} R.,  {Beaumont} C.~N.,  {Burton} M.~G.,  {Kelly} B.~C.,   {Klessen}
  R.~S.,  2012, \mnras, 425, 720

\bibitem[\protect\citeauthoryear{{Sormani} \& {Barnes}}{{Sormani} \&
  {Barnes}}{2019}]{sormani19}
{Sormani} M.~C.,  {Barnes} A.~T.,  2019, \mn@doi [\mnras]
  {10.1093/mnras/stz046}, \href
  {https://ui.adsabs.harvard.edu/abs/2019MNRAS.484.1213S} {484, 1213}

\bibitem[\protect\citeauthoryear{{Sormani}, {Binney}  \& {Magorrian}}{{Sormani}
  et~al.}{2015}]{sormani15}
{Sormani} M.~C.,  {Binney} J.,   {Magorrian} J.,  2015, \mn@doi [\mnras]
  {10.1093/mnras/stv441}, \href
  {https://ui.adsabs.harvard.edu/abs/2015MNRAS.449.2421S} {449, 2421}

\bibitem[\protect\citeauthoryear{{Tang}, {Wang}  \& {Wilson}}{{Tang}
  et~al.}{2020}]{tang20}
{Tang} Y.,  {Wang} Q.~D.,   {Wilson} G.~W.,  2020, arXiv e-prints, p.
  arXiv:2008.12361

\bibitem[\protect\citeauthoryear{{Tsuboi}, {Miyazaki}  \& {Uehara}}{{Tsuboi}
  et~al.}{2015}]{tsuboi15}
{Tsuboi} M.,  {Miyazaki} A.,   {Uehara} K.,  2015, \mn@doi [\pasj]
  {10.1093/pasj/psv076}, \href
  {https://ui.adsabs.harvard.edu/abs/2015PASJ...67..109T} {67, 109}

\bibitem[\protect\citeauthoryear{{Tsuboi}, {Kitamura}, {Uehara}, {Miyawaki},
  {Tsutsumi}, {Miyazaki}  \& {Miyoshi}}{{Tsuboi} et~al.}{2021}]{tsuboi21}
{Tsuboi} M.,  {Kitamura} Y.,  {Uehara} K.,  {Miyawaki} R.,  {Tsutsumi} T.,
  {Miyazaki} A.,   {Miyoshi} M.,  2021, \mn@doi [\pasj] {10.1093/pasj/psaa095},
  \href {https://ui.adsabs.harvard.edu/abs/2021PASJ...73S..91T} {73, S91}

\bibitem[\protect\citeauthoryear{{Urquhart} et~al.,}{{Urquhart}
  et~al.}{2013}]{urquhart13}
{Urquhart} J.~S.,  et~al., 2013, \mn@doi [\mnras] {10.1093/mnras/stt287}, \href
  {https://ui.adsabs.harvard.edu/abs/2013MNRAS.431.1752U} {431, 1752}

\bibitem[\protect\citeauthoryear{{Walker}, Longmore, Bastian, Kruijssen,
  Rathborne, Jackson, Foster  \& Contreras}{{Walker} et~al.}{2015}]{walker15}
{Walker} D.~L.,  Longmore S.~N.,  Bastian N.,  Kruijssen J. M.~D.,  Rathborne
  J.~M.,  Jackson J.~M.,  Foster J.~B.,   Contreras Y.,  2015, \mnras, 449, 715

\bibitem[\protect\citeauthoryear{{Walker}, {Longmore}, {Bastian}, {Kruijssen},
  {Rathborne}, {Galv{\'a}n-Madrid}  \& {Liu}}{{Walker} et~al.}{2016}]{walker16}
{Walker} D.~L.,  {Longmore} S.~N.,  {Bastian} N.,  {Kruijssen} J.~M.~D.,
  {Rathborne} J.~M.,  {Galv{\'a}n-Madrid} R.,   {Liu} H.~B.,  2016, \mnras,
  457, 4536

\bibitem[\protect\citeauthoryear{{Walker} et~al.,}{{Walker}
  et~al.}{2018}]{walker18}
{Walker} D.~L.,  et~al., 2018, \mnras, 474, 2373

\bibitem[\protect\citeauthoryear{{Walker} et~al.,}{{Walker}
  et~al.}{2021}]{walker21}
{Walker} D.~L.,  et~al., 2021, \mnras, 503, 77

\bibitem[\protect\citeauthoryear{{Zeng} et~al.,}{{Zeng} et~al.}{2020}]{zeng20}
{Zeng} S.,  et~al., 2020, \mn@doi [\mnras] {10.1093/mnras/staa2187}, \href
  {https://ui.adsabs.harvard.edu/abs/2020MNRAS.497.4896Z} {497, 4896}

\makeatother
\end{thebibliography}



\appendix

\section{Tables}

\begin{table*}
	\centering
	\caption{\emph{Note: this is a sample table, the full table is available online as supplementary material. Masses, radii and number densities (assuming spherical geometry) of each the 96 individual potential star-forming sources} we have detected using dendrogram analysis, as well as estimated masses and number densities with background emission subtracted.}
	\label{tab:dendfull}
	\begin{tabular}{cccccccc} 
		\hline
		\hline
		 & Right & & & & & Background & Background \\
		 Source Number & Ascension & Declination & Mass & Radius & n & subtracted mass & subtracted n\\
		 & (degrees) & (degrees) & (M${}_{\odot}$) & (10${}^3$\,au) & (10${}^6$ cm${}^{-3}$) & (M${}_{\odot}$) & (10${}^6$ cm${}^{-3}$)\\
		\hline
        1  & 266.5949552 & -28.56055698 & 0.71 & 1.71 & 6.04  & 0.06 & 0.54 \\
        2  & 266.5950498 & -28.56040786 & 2.82 & 3.31 & 3.3   & 0.39 & 0.45 \\
        3  & 266.5921428 & -28.56046861 & 0.92 & 1.96 & 5.21  & 0.23 & 1.28 \\
        4  & 266.5954069 & -28.56032063 & 1.00    & 2.02 & 5.14  & 0.22 & 1.14 \\
        5  & 266.5947927 & -28.56023752 & 0.48 & 1.42 & 7.06  & 0.05 & 0.69 \\
        6  & 266.5949923 & -28.56017972 & 0.25 & 1.04 & 9.42  & 0.04 & 1.32 \\
        7  & 266.5948637 & -28.56005386 & 0.76 & 1.78 & 5.67  & 0.08 & 0.62 \\
        8  & 266.5947918 & -28.55997624 & 0.4  & 1.31 & 7.57  & 0.05 & 0.95 \\
        9  & 266.594819  & -28.55975369 & 0.35 & 1.3  & 6.7   & 0.04 & 0.78 \\
        10 & 266.5947906 & -28.55961351 & 0.37 & 1.33 & 6.61  & 0.04 & 0.81 \\
        \hline
	\end{tabular}
\end{table*}

\begin{table*}
	\centering
	\caption{Virial parameter, $\alpha$, and velocity dispersion, $\sigma$, (and associated upper and lower limits) of sources for which velocity dispersion could be measured in H${}_2$CO (3${}_{0,3}$ - 2${}_{0,2}$). Asterisks denote the sources found using a 5$\sigma$ threshold.}
	\label{tab:virial}
	\begin{tabular}{cccccc} 
		\hline
		\hline
        Source & $\alpha$ & $\alpha_{lower}$ & $\alpha_{\mathrm upper}$ & $\sigma$ & $\Delta\sigma$\\
        Number & H${}_2$CO (3${}_{0,3}$ - 2${}_{0,2}$) & & & kms${}^{-1}$ & kms${}^{-1}$\\
        \hline
        19 & 8.353476  & 2.099979  & 18.760553  & 0.752865 & 0.375388 \\
        22 & 27.004411 & 2.600424  & 77.09842   & 1.023652 & 0.705996 \\
        29 & 10.909824 & 0.106466  & 39.434801  & 1.030012 & 0.928261 \\
        30 & 10.468572 & 3.335144  & 21.574139  & 1.055967 & 0.459943 \\
        50 & 44.978999 & 11.594935 & 100.162932 & 1.289734 & 0.634902 \\
        55 & 12.351154 & 0.910142  & 36.903537  & 0.761981 & 0.555136 \\
        56 & 5.417488  & 0.81414   & 14.083529  & 0.586945 & 0.35941  \\
        58 & 25.066655 & 7.456437  & 53.037347  & 1.419999 & 0.645527 \\
        79 & 29.780918 & 5.930561  & 71.895217  & 1.199221 & 0.664068 \\
        80* & 6.147653  & 1.128015  & 15.185141  & 0.846895 & 0.484124 \\
        81 & 22.668    & 5.600116  & 51.204446  & 0.923663 & 0.464565 \\
        82 & 5.119289  & 0.055064  & 18.408492  & 0.732269 & 0.656324 \\
        85 & 12.289794 & 4.493792  & 23.926809  & 0.914435 & 0.361484 \\
        \hline
	\end{tabular}
\end{table*}
\bsp	
\label{lastpage}
\end{document}


\pagenumbering{gobble}

\begin{figure*}
	    \begin{center}
        \includegraphics[width=16cm]{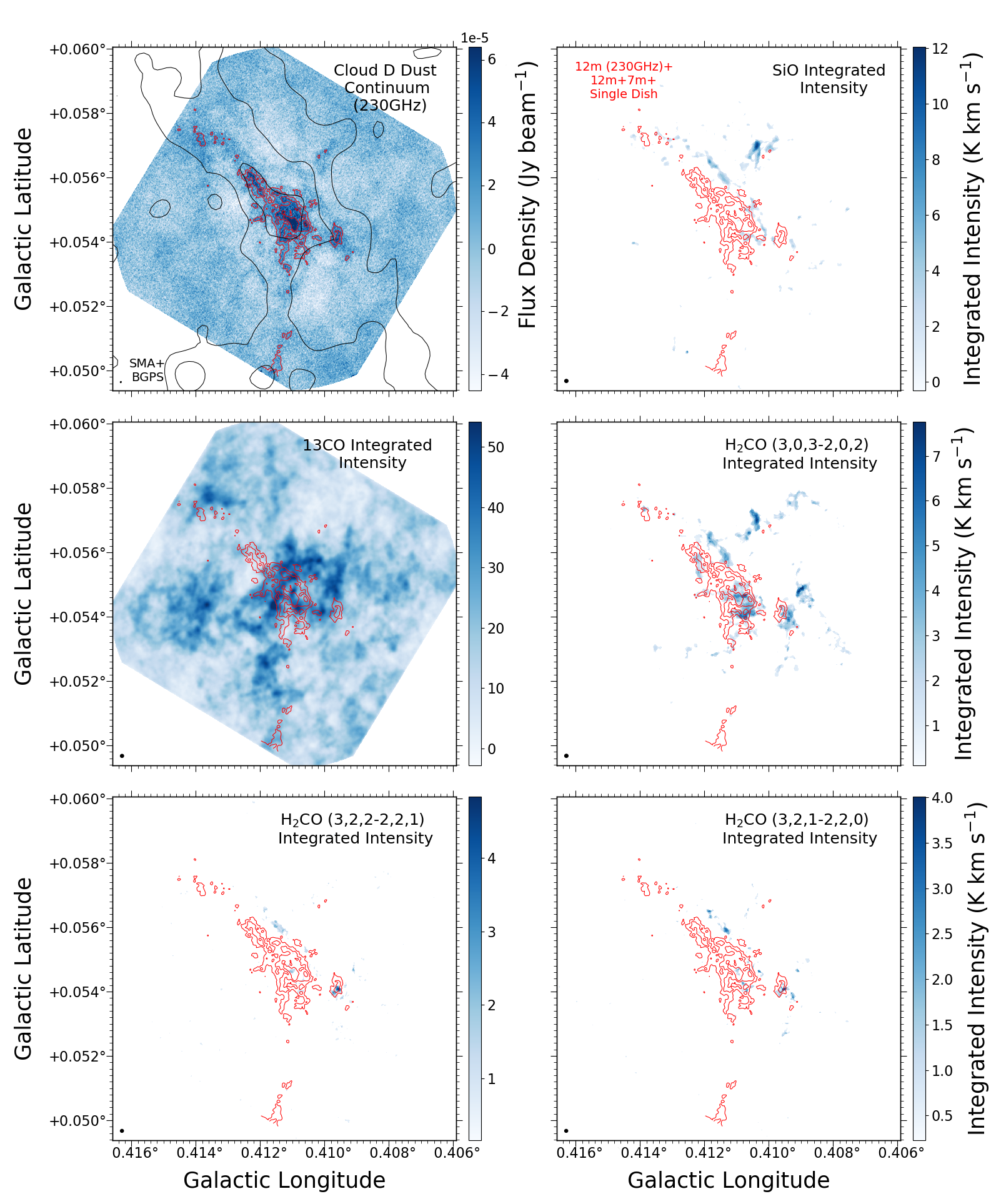}
		\caption{Upper left: 230 GHz dust continuum. All other plots: moment 0 (integrated intensity) maps of cloud `d' in various spectral lines, with comparison to dust continuum. Black contours $=$ SMA$+$BGPS data. Red contours = dust continuum.}
		\label{fig:mom0}
		\end{center}
    \end{figure*}
    
\begin{figure*}
	    \begin{center}
        \includegraphics[width=16cm]{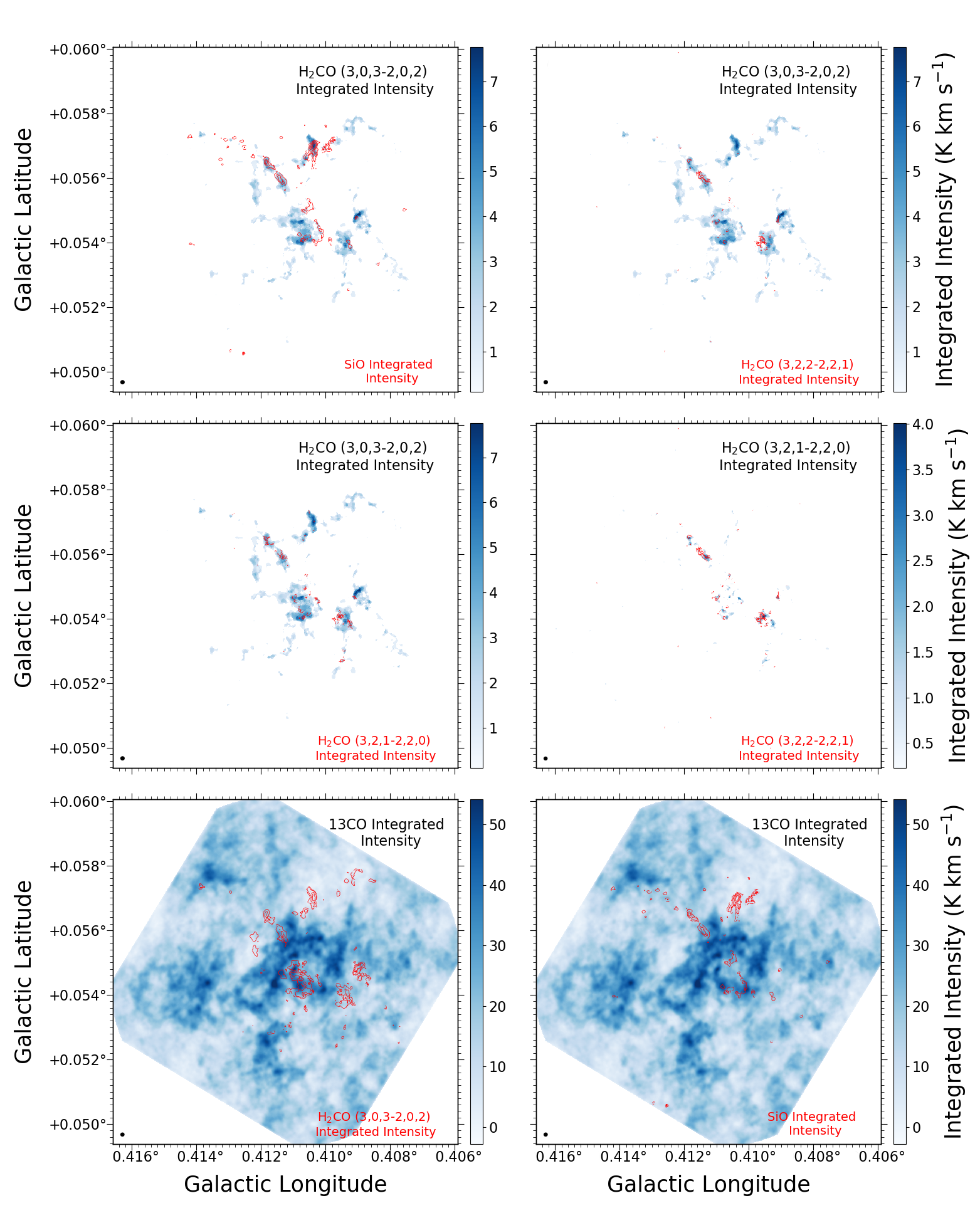}
		\caption{Moment 0 (integrated intensity) maps of cloud `d' in various spectral lines, with comparison to other spectral lines. Black contours $=$ SMA$+$BGPS data. Red contours $=$ spectral line for comparison (see individual plots for details).}
		\label{fig:mom0comp}
		\end{center}
    \end{figure*}
    
\begin{figure*}
	    \begin{center}
        \includegraphics[width=16cm]{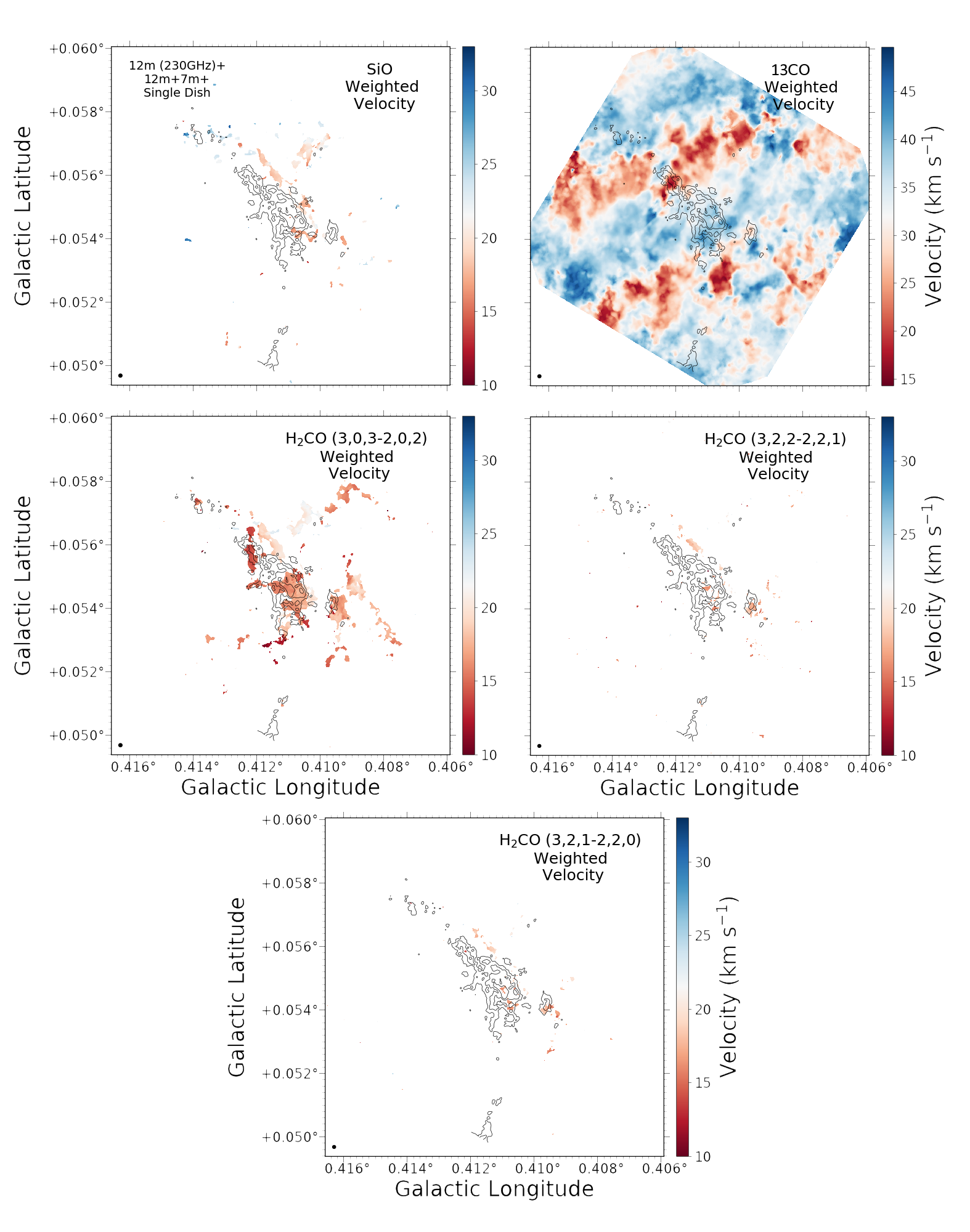}
		\caption{Moment 1 (weighted velocity) maps of cloud `d' in various spectral lines, with comparison to dust continuum. Black contours = dust continuum.}
		\label{fig:mom1}
		\end{center}
    \end{figure*}

\begin{figure*}
	    \begin{center}
        \includegraphics[width=16cm]{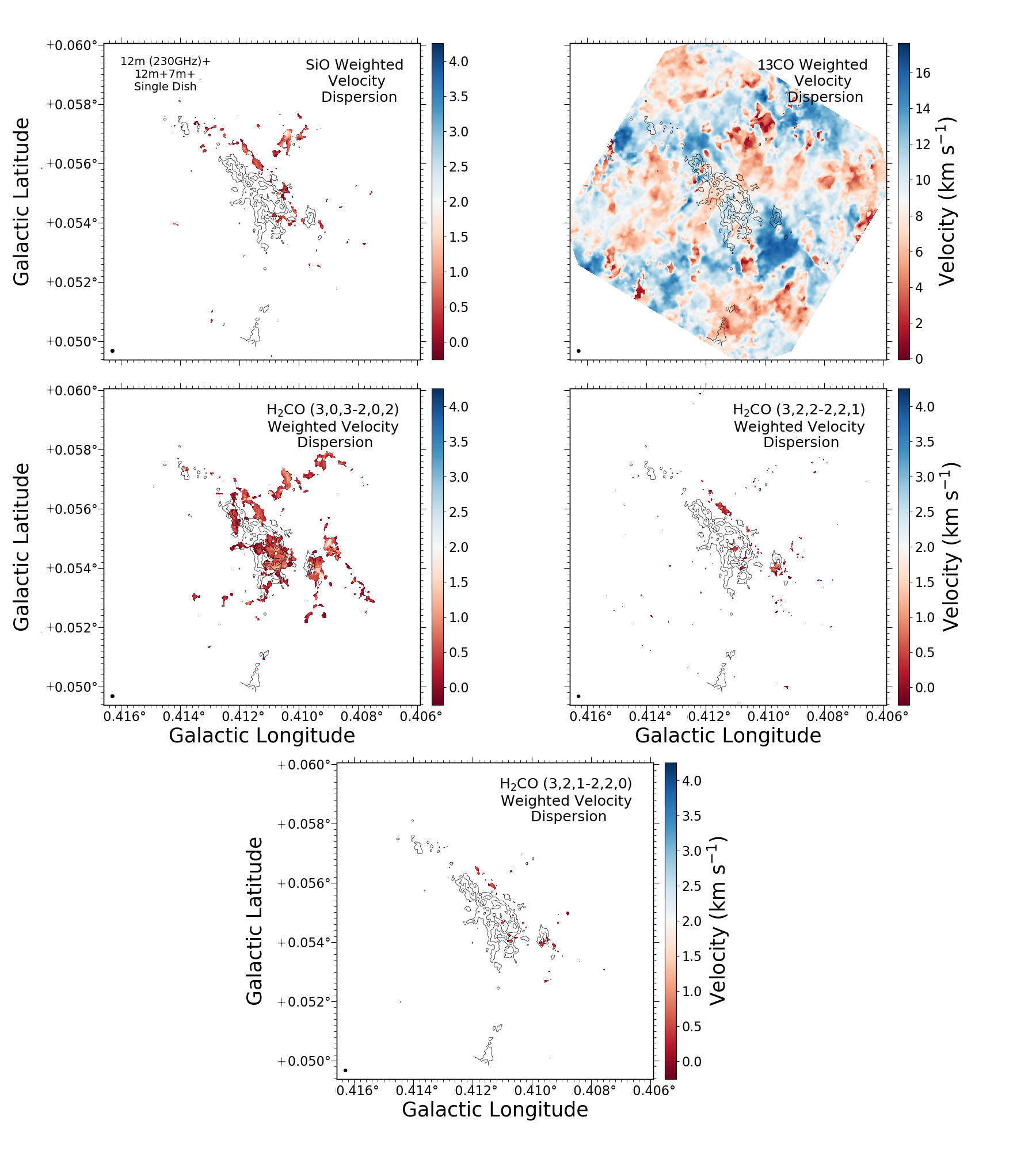}
		\caption{Moment 2 (weighted velocity dispersion) maps of cloud `d' in various spectral lines, with comparison to dust continuum. Black contours = dust continuum.}
		\label{fig:mom2}
		\end{center}
    \end{figure*}


\pagenumbering{gobble}

\begin{figure*}
	    \begin{center}
        \includegraphics[width=16cm]{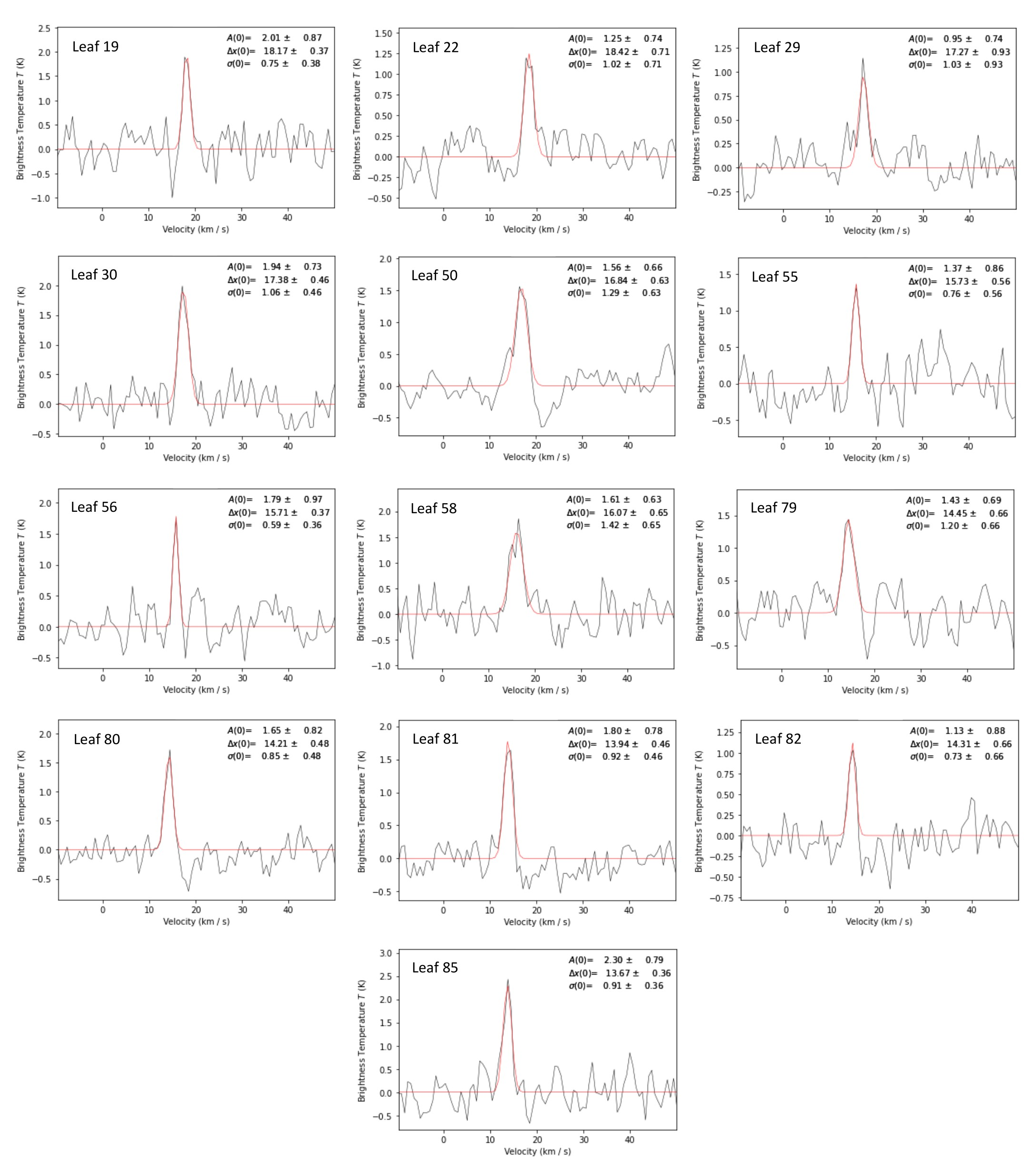}
		\caption{Gaussian fits of H${}_2$CO (3${}_{0,3}$ - 2${}_{0,2}$) spectra towards all dendrogram leaves with a clear line detection. The upper right corner displays the amplitude of the line at the central velocity, the central velocity, and the velocity dispersion, along with errors for each of these values.}
		\label{fig:spect}
		\end{center}
    \end{figure*}


\pagenumbering{gobble}

\begin{table*}
	\centering
	\caption{Masses, radii and number densities (assuming spherical geometry) of each the 96 individual potential star-forming sources we have detected using dendrogram analysis, as well as estimated masses and number densities with background emission subtracted. The sources are all low-mass, with the most massive being 3.08 M${}_{\odot}$. Asterisks denote the nine sources found using a 5$\sigma$ threshold.}
	\label{tab:dendfull}
	\begin{tabular}{cccccccc} 
		\hline
		\hline
		 & Right & & & & & Background & Background \\
		 Source Number & Ascension & Declination & Mass & Radius & n & subtracted mass & subtracted n\\
		 & (degrees) & (degrees) & (M${}_{\odot}$) & (10${}^3$\,au) & (10${}^6$ cm${}^{-3}$) & (M${}_{\odot}$) & (10${}^6$ cm${}^{-3}$)\\
		\hline
        1  & 266.5949552 & -28.56055698 & 0.71 & 1.71 & 6.04  & 0.06 & 0.54 \\
        2  & 266.5950498 & -28.56040786 & 2.82 & 3.31 & 3.3   & 0.39 & 0.45 \\
        3  & 266.5921428 & -28.56046861 & 0.92 & 1.96 & 5.21  & 0.23 & 1.28 \\
        4  & 266.5954069 & -28.56032063 & 1    & 2.02 & 5.14  & 0.22 & 1.14 \\
        5  & 266.5947927 & -28.56023752 & 0.48 & 1.42 & 7.06  & 0.05 & 0.69 \\
        6  & 266.5949923 & -28.56017972 & 0.25 & 1.04 & 9.42  & 0.04 & 1.32 \\
        7  & 266.5948637 & -28.56005386 & 0.76 & 1.78 & 5.67  & 0.08 & 0.62 \\
        8  & 266.5947918 & -28.55997624 & 0.4  & 1.31 & 7.57  & 0.05 & 0.95 \\
        9  & 266.594819  & -28.55975369 & 0.35 & 1.3  & 6.7   & 0.04 & 0.78 \\
        10 & 266.5947906 & -28.55961351 & 0.37 & 1.33 & 6.61  & 0.04 & 0.81 \\
        11 & 266.5946396 & -28.55924377 & 0.4  & 1.41 & 5.98  & 0.04 & 0.54 \\
        12 & 266.5944071 & -28.55924543 & 0.32 & 1.24 & 7.09  & 0.04 & 0.87 \\
        13 & 266.594379  & -28.55913663 & 0.24 & 1.07 & 8.31  & 0.03 & 1.08 \\
        14 & 266.5944861 & -28.55789014 & 0.24 & 1.05 & 8.72  & 0.04 & 1.35 \\
        15 & 266.5944708 & -28.55723895 & 0.26 & 1.09 & 8.56  & 0.03 & 0.84 \\
        16 & 266.594455  & -28.557144   & 0.4  & 1.32 & 7.31  & 0.04 & 0.82 \\
        17 & 266.5945802 & -28.5571069  & 0.55 & 1.53 & 6.49  & 0.07 & 0.88 \\
        18 & 266.5938228 & -28.55699922 & 0.35 & 1.32 & 6.44  & 0.03 & 0.61 \\
        19 & 266.5947139 & -28.55691578 & 0.65 & 1.7  & 5.58  & 0.08 & 0.67 \\
        20 & 266.5923978 & -28.55684902 & 0.67 & 1.81 & 4.81  & 0.08 & 0.55 \\
        21 & 266.5939338 & -28.5568055  & 0.4  & 1.38 & 6.46  & 0.05 & 0.87 \\
        22 & 266.5942263 & -28.55683809 & 0.23 & 1.05 & 8.44  & 0.03 & 1.06 \\
        23 & 266.5941399 & -28.55677423 & 0.8  & 1.96 & 4.48  & 0.1  & 0.57 \\
        24 & 266.594711  & -28.55678025 & 0.4  & 1.29 & 7.77  & 0.05 & 0.9  \\
        25 & 266.5946677 & -28.55669591 & 0.36 & 1.21 & 8.66  & 0.05 & 1.32 \\
        26 & 266.5938146 & -28.55665411 & 0.83 & 1.89 & 5.21  & 0.14 & 0.85 \\
        27 & 266.5938906 & -28.55652185 & 0.44 & 1.43 & 6.38  & 0.04 & 0.63 \\
        28 & 266.5928481 & -28.55642795 & 0.45 & 1.39 & 7.11  & 0.04 & 0.6  \\
        29 & 266.5940068 & -28.55637764 & 1.27 & 2.32 & 4.3   & 0.22 & 0.75 \\
        30 & 266.5927609 & -28.55627968 & 1.32 & 2.2  & 5.23  & 0.17 & 0.65 \\
        31 & 266.5947371 & -28.55626461 & 0.4  & 1.19 & 10.01 & 0.04 & 1.05 \\
        32 & 266.5945867 & -28.55627705 & 0.61 & 1.5  & 7.71  & 0.05 & 0.59 \\
        33 & 266.5934777 & -28.55621856 & 1.26 & 2.45 & 3.65  & 0.17 & 0.5  \\
        34 & 266.5941926 & -28.55619617 & 0.36 & 1.17 & 9.74  & 0.03 & 0.91 \\
        35 & 266.5929402 & -28.55621515 & 0.29 & 1.15 & 8.09  & 0.04 & 1.12 \\
        36 & 266.594683  & -28.556048   & 1.45 & 2.27 & 5.22  & 0.16 & 0.57 \\
        37 & 266.5943944 & -28.55614844 & 0.94 & 1.87 & 6.02  & 0.08 & 0.54 \\
        38 & 266.5926697 & -28.55614251 & 0.36 & 1.15 & 10.15 & 0.05 & 1.36 \\
        39 & 266.5928167 & -28.55603091 & 0.43 & 1.31 & 8.06  & 0.03 & 0.64 \\
        40 & 266.5926899 & -28.55603956 & 0.39 & 1.23 & 9     & 0.04 & 0.93 \\
        41 & 266.5935172 & -28.55596575 & 0.36 & 1.33 & 6.48  & 0.04 & 0.71 \\
        42 & 266.5938055 & -28.55597085 & 0.29 & 1    & 12.23 & 0.03 & 1.21 \\
        43* & 266.5942592 & -28.55591566 & 0.87 & 1.44 & 12.38 & 0.06 & 0.92 \\
        44* & 266.5938724 & -28.55582407 & 3.08 & 3.12 & 4.28  & 0.43 & 0.6  \\
        45 & 266.5944476 & -28.55589948 & 0.58 & 1.46 & 7.85  & 0.05 & 0.66 \\
        46 & 266.5947276 & -28.55579846 & 0.4  & 1.3  & 7.74  & 0.05 & 0.93 \\
        47 & 266.5945846 & -28.55584819 & 0.37 & 1.18 & 9.75  & 0.03 & 0.77 \\
        48 & 266.5948484 & -28.55577893 & 0.58 & 1.53 & 6.8   & 0.05 & 0.58 \\
        49* & 266.5942739 & -28.55573905 & 1.81 & 2.06 & 8.77  & 0.16 & 0.78 \\
        50 & 266.5957627 & -28.55575829 & 0.21 & 1.01 & 8.66  & 0.02 & 1.02 \\
        51 & 266.5948947 & -28.55566816 & 0.74 & 1.72 & 6.12  & 0.07 & 0.57 \\
        52 & 266.5954151 & -28.55563759 & 0.9  & 1.92 & 5.4   & 0.14 & 0.83 \\
        53 & 266.5956096 & -28.55564977 & 0.44 & 1.36 & 7.42  & 0.06 & 0.97 \\
        54* & 266.5946004 & -28.55558255 & 0.76 & 1.51 & 9.4   & 0.07 & 0.85 \\
        55 & 266.5952839 & -28.55550218 & 0.32 & 1.19 & 7.88  & 0.05 & 1.29 \\
        56 & 266.5948508 & -28.55548764 & 0.4  & 1.11 & 12.36 & 0.05 & 1.62 \\
        57* & 266.5946633 & -28.5554291  & 1.36 & 2.05 & 6.7   & 0.18 & 0.9  \\
        58 & 266.5950172 & -28.55537882 & 0.84 & 1.85 & 5.59  & 0.1  & 0.66 \\
        \hline
	\end{tabular}
\end{table*}

\begin{table*}
	\centering
	\caption{Masses, radii and number densities (assuming spherical geometry) of each the 96 individual potential star-forming sources we have detected using dendrogram analysis, as well as estimated masses and number densities with background emission subtracted. The sources are all low-mass, with the most massive being 3.08 M${}_{\odot}$. Asterisks denote the nine sources found using a 5$\sigma$ threshold.}
	\label{tab:dendfull2}
	\begin{tabular}{cccccccc} 
		\hline
		\hline
		 & Right & & & & & Background & Background \\
		Source Number & Ascension & Declination & Mass & Radius & n & subtracted mass & subtracted n\\
		 & (degrees) & (degrees) & (M${}_{\odot}$) & (10${}^3$\,au) & (10${}^6$ cm${}^{-3}$) & (M${}_{\odot}$) & (10${}^6$ cm${}^{-3}$)\\
		\hline
		59 & 266.5945059 & -28.55535908 & 0.67 & 1.47 & 8.84  & 0.06 & 0.84 \\
        60* & 266.59428   & -28.55533156 & 0.84 & 1.63 & 8.17  & 0.07 & 0.69 \\
        61* & 266.5941472 & -28.55534326 & 0.64 & 1.39 & 10.16 & 0.09 & 1.4  \\
        62 & 266.5952992 & -28.55528494 & 0.27 & 1.13 & 8.06  & 0.03 & 0.9  \\
        63 & 266.594749  & -28.55528651 & 0.33 & 1.08 & 11.02 & 0.03 & 1    \\
        64 & 266.5951921 & -28.55523682 & 0.49 & 1.49 & 6.31  & 0.06 & 0.77 \\
        65 & 266.5956213 & -28.55523389 & 0.71 & 1.82 & 4.97  & 0.09 & 0.64 \\
        66 & 266.5945798 & -28.55527245 & 0.34 & 1.08 & 11.17 & 0.03 & 1.07 \\
        67 & 266.5936359 & -28.55518    & 0.42 & 1.44 & 5.93  & 0.04 & 0.59 \\
        68 & 266.595738  & -28.55519265 & 0.32 & 1.23 & 7.45  & 0.04 & 1    \\
        69 & 266.5948717 & -28.55509246 & 0.5  & 1.44 & 7.03  & 0.06 & 0.79 \\
        70 & 266.5937017 & -28.55511106 & 0.26 & 1.1  & 8.08  & 0.03 & 1.08 \\
        71 & 266.595262  & -28.55505473 & 0.51 & 1.55 & 5.88  & 0.05 & 0.62 \\
        72 & 266.594601  & -28.55503243 & 1.49 & 2.31 & 5.08  & 0.17 & 0.58 \\
        73 & 266.5944452 & -28.5550415  & 0.28 & 1    & 11.59 & 0.03 & 1.21 \\
        74 & 266.5951052 & -28.55501255 & 0.46 & 1.52 & 5.64  & 0.05 & 0.56 \\
        75 & 266.595397  & -28.55486812 & 0.99 & 1.9  & 6.08  & 0.11 & 0.7  \\
        76 & 266.5949318 & -28.55482544 & 1.77 & 2.62 & 4.17  & 0.21 & 0.49 \\
        77 & 266.5941558 & -28.55481454 & 1.2  & 2.35 & 3.9   & 0.19 & 0.62 \\
        78 & 266.5947667 & -28.55483111 & 0.27 & 1.05 & 10.07 & 0.02 & 0.88 \\
        79 & 266.5957172 & -28.55474527 & 0.34 & 1.24 & 7.48  & 0.04 & 0.87 \\
        80* & 266.595471  & -28.5547302  & 1.4  & 2.13 & 6.11  & 0.22 & 0.98 \\
        81 & 266.595185  & -28.5547352  & 0.22 & 1.03 & 8.39  & 0.03 & 0.96 \\
        82 & 266.5955859 & -28.5545708  & 1.19 & 2.01 & 6.19  & 0.14 & 0.73 \\
        83 & 266.5958693 & -28.55458649 & 0.42 & 1.25 & 9.15  & 0.05 & 1.08 \\
        84* & 266.59577   & -28.55432102 & 2.76 & 3.07 & 4.03  & 0.39 & 0.58 \\
        85 & 266.5953755 & -28.55447055 & 0.59 & 1.55 & 6.79  & 0.07 & 0.81 \\
        86 & 266.5955486 & -28.55439258 & 0.36 & 1.19 & 9.17  & 0.05 & 1.27 \\
        87 & 266.5962215 & -28.55419811 & 0.64 & 1.76 & 4.97  & 0.08 & 0.59 \\
        88 & 266.5962714 & -28.55367059 & 0.43 & 1.45 & 6.02  & 0.05 & 0.72 \\
        89 & 266.5975016 & -28.55328759 & 0.87 & 2.03 & 4.4   & 0.11 & 0.56 \\
        90 & 266.5967906 & -28.55329433 & 0.29 & 1.19 & 7.2   & 0.03 & 0.75 \\
        91 & 266.5970503 & -28.55316031 & 0.6  & 1.72 & 4.99  & 0.06 & 0.51 \\
        92 & 266.597619  & -28.55314851 & 0.83 & 2    & 4.43  & 0.09 & 0.46 \\
        93 & 266.5975601 & -28.55301651 & 0.26 & 1.13 & 7.83  & 0.03 & 0.76 \\
        94 & 266.5977636 & -28.55284698 & 0.52 & 1.58 & 5.59  & 0.07 & 0.7  \\
        95 & 266.5983306 & -28.55286838 & 0.27 & 1.14 & 7.66  & 0.03 & 0.93 \\
        96 & 266.5983951 & -28.55269641 & 0.44 & 1.45 & 6.06  & 0.06 & 0.77 \\
		\hline
	\end{tabular}
\end{table*}